\documentclass[prb,twocolumn,showpacs,floatfix,balancelastpage]{revtex4-1}
\usepackage[latin1]{inputenc}
\usepackage{graphicx}
\usepackage{amssymb}
\usepackage{amsmath}
\usepackage{xspace}
\usepackage{dcolumn}
\usepackage{bm}
\usepackage{color}

\usepackage[extra]{tipa}

\setcitestyle{square,numbers}

\newfont{\tensy}{cmsy10}

\newcommand{\ie}[0]{i.e.\@\xspace}

\newcommand{\UP}[0]{\uparrow}
\newcommand{\DO}[0]{\downarrow}

\newcommand{\on}{\hat{n}}

\newcommand{\bit}{\begin{itemize}}
\newcommand{\eit}{\end{itemize}}

\newcommand{\om}[0]{\omega}

\newcommand{\kF}{k_\text{F}}

\newcommand{\nag}{{\phantom{\dag}}}

\newcommand{\las}[0]{\langle}
\newcommand{\ras}[0]{\rangle}

\renewcommand{\tilde}[1]{\widetilde{#1}}

\begin{document}


\title{Phonon spectral function of the one-dimensional Holstein-Hubbard model}

\author{Manuel Weber}

\author{Fakher F. Assaad}

\author{Martin Hohenadler}

\affiliation{\mbox{Institut f\"ur Theoretische Physik und Astrophysik,
    Universit\"at W\"urzburg, 97074 W\"urzburg, Germany}}

\begin{abstract}
  We use the continuous-time interaction expansion (CT-INT) quantum Monte
  Carlo method to calculate the phonon spectral function of the
  one-dimensional Holstein-Hubbard model at half-filling. Our results are
  consistent with a soft-mode Peierls transition in the adiabatic regime, and
  the existence of a central peak related to long-range order in the Peierls
  phase.  We explain a previously observed feature at small momenta in terms
  of a hybridization of charge and phonon excitations.  Tuning the system
  from a Peierls to a metallic phase with a nonzero Hubbard interaction
  suppresses the central peak, but a significant renormalization of the
  phonon dispersion remains. In contrast, the
  dispersion is only weakly modified in the Mott phase. We discuss finite-size effects, the relation to
  the dynamic charge structure factor, as well as additional sum rules and
  their implications. Finally, we reveal the existence of a discrete symmetry
  in a continuum field theory of the Holstein model, which is spontaneously
  broken in the Peierls phase.
\end{abstract}

\date{\today}

\pacs{71.10.Pm, 71.45.Lr, 71.30.+h, 71.38.-k} 


\maketitle

\section{Introduction}\label{sec:intro}

Electron-phonon interaction plays a crucial role for the physics of materials
\cite{polaronbook2007}, and gives rise to phenomena such as the Peierls
instability \cite{Peierls}, superconductivity, and polaron formation
\cite{0034-4885-72-6-066501}. Recently, interest in electron-phonon
interaction has also been boosted by experiments involving photo-induced
phase transitions between insulating Peierls and metallic states
\cite{Chollet07012005}, see Ref.~\cite{Yonemitsu20081} for a review.

The study of microscopic electron-phonon models has a long and rich
history. Much of the recent progress on the theoretical side resulted from
the development of exact numerical methods, most notably the density-matrix
renormalization group \cite{Schollwoeck05_rev,BuMKHa98,JeZhWh99}, and
quantum Monte Carlo (QMC) methods
\cite{BlScSu81,dRLa82,PhysRevB.27.1680,Ko98,PrSv98,PhysRevLett.83.195,Assaad07}. Whereas
the case of a single electron (the polaron problem) can be solved to machine
precision using a variational basis construction \cite{BoTrBa99}, finite band
fillings are still a challenge.  Here, we discuss one-dimensional (1D)
models.

Of particular interest with regard to experiment is the calculation of
spectral properties such as the electronic spectral function that may be
compared to angular-resolved photoemission spectra. For the spinless Holstein
model at half-filling, it reveals the opening of a Peierls gap for strong
electron-phonon coupling \cite{SyHuBeWeFe04,ZhWuLi05,Hohenadler06}, the
experimentally observed shadow bands \cite{Vo.Pe.Zw.Be.Ma.Gr.Ho.Gr.00}
arising from the new periodicity of the lattice \cite{Hohenadler10a}, and
soliton excitations \cite{Hohenadler10a}.  The phonon spectral function and
the dynamic charge structure factor---experimentally accessible via neutron
scattering---reveal, for example, the softening of phonon excitations near
the Peierls transition \cite{CrSaCa05,Hohenadler06,Sykora06}, and distinguish
soft-mode behavior in the adiabatic regime from central-peak behavior in the
nonadiabatic regime.  For a single electron, the phonon spectrum can again be
calculated to arbitrary precision \cite{LoHoAlFe06}. The phonon spectral
function and the renormalized phonon frequency of the spinless Holstein model
were obtained with the projector-based renormalization method
\cite{SyHuBeWeFe04,SyHuBe05,Sykora06}.  The electronic spectral function of
the Holstein-Hubbard model was calculated to characterize the metallic, Mott,
and Peierls phases \cite{FeWeHaWeBi03,PhysRevB.83.033104,PhysRevB87.075149},
and to study the impact of electron-phonon coupling on spin-charge separation
\cite{NiZhWuLi05,MaToMa05}. A more complete review of previous work can be
found in Refs.~\cite{Hohenadler10a,PhysRevB87.075149}.

Here, we show that the CT-INT method, which was applied to a number of
electron-phonon problems
\cite{Assaad07,Assaad08,Hohenadler10a,Ho.As.Fe.12,PhysRevB87.075149,PhysRevB.88.064303,WeAsHo15_I},
can also be used to calculate the phonon Green function, and thereby the phonon
spectral function. The method is free of a Trotter error, and does not
require a cutoff for the phonon Hilbert space. We use it to calculate the
phonon spectral function of the half-filled Holstein-Hubbard model. The
results are discussed in the context of previous work on these and other
models, and additional insights are provided with the help of the random
phase approximation and field theory.

The paper is organized as follows. In Sec.~\ref{sec:model} we define the
model. In Sec.~\ref{sec:method} we discuss the method. Our results are
presented in Sec.~\ref{sec:results}, followed by a discussion in
Sec.~\ref{sec:discussion}. Section~\ref{sec:conclusions} contains
our conclusions, and the Appendix discusses the relation between
charge and phonon spectra, sum rules, and their implications.

\section{Model}\label{sec:model}

We consider the Holstein-Hubbard model
\begin{align}\label{eq:model-holsteinspinful}
  \hat{H}  
  &=
  -t\sum_{i\sigma} \left( \hat{c}^{\dag}_{i\sigma} \hat{c}^\nag_{i+1\sigma} + \text{H.c.} \right)
 + U \sum_i \on_{i\UP} \on_{i\DO}
  \\\nonumber
  &\quad+ \sum_{i} \left(\mbox{$\frac{1}{2M}$} \hat{P}_{i}^2 + \mbox{$\frac{K}{2}$}
    \hat{Q}_{i}^2 \right)
  - g \sum_{i} \hat{Q}_{i} \hat{\rho}_{i}
  \,.
\end{align}
The first term describes the hopping of electrons between neighboring lattice
sites with amplitude $t$.  The second term captures the repulsion between
electrons at the same lattice site; for $U=0$,
Hamiltonian~(\ref{eq:model-holsteinspinful}) becomes the spinful Holstein
model \cite{Ho59a}. The lattice is described in the harmonic approximation, with the
displacement (momentum) at site $i$ given by $\hat{Q}_i$ ($\hat{P}_i$). The
bare optical phonon frequency is $\om_0=\sqrt{K/M}$. The electron-phonon
interaction is of the density-displacement type, with $\hat{\rho}_i =
\on_i-1$, $\hat{n}_i = \sum_\sigma \hat{n}_{i\sigma}$, and $\hat{n}_{i\sigma}
= \hat{c}^{\dag}_{i\sigma } \hat{c}^\nag_{i\sigma }$.  The dimensionless
ratio $\lambda=g^2/(4Kt)$ is a useful measure of the electron-phonon
coupling strength.

We studied Eq.~(\ref{eq:model-holsteinspinful}) at half-filling,
corresponding to $\las\hat{n}_i\ras=1$, on chains with $L$ sites and with
periodic boundary conditions. We use $t$ as the unit of energy, and set the
lattice constant, $M$, and $\hbar$ to one.

The Holstein-Hubbard model captures the Peierls transition driven by strong
electron-phonon coupling, the Mott physics related to strong
electron-electron interaction, and the competition between these two
phenomena when $U\sim \lambda$
\cite{FeWeHaWeBi03,ClHa05,hardikar:245103,0295-5075-84-5-57001,Bakrim2015}. Although
some open questions remain regarding the phase diagram and the low-energy
physics, the prevailing picture is that the model supports an intermediate
metallic phase with a spin gap but gapless density fluctuations
\cite{PhysRevB87.075149,Bakrim2015}, which is adiabatically connected to the
metallic phase of the spinful Holstein model ($U=0$) \cite{JeZhWh99}. The
metallic behavior for $\lambda<\lambda_{c}(U,\om_0)$ is the result of quantum
lattice fluctuations that destroy the long-range Peierls order. Its extent
therefore depends on the phonon frequency: For $\om_0=0$ a Peierls state
exists as soon as $4\lambda t>U$, whereas for $\om_0=\infty$ a Peierls state
is completely absent from the phase diagram
\cite{ClHa05,hardikar:245103,0295-5075-84-5-57001,Bakrim2015}. In the
classical limit and for half-filling, the charge order in the Peierls phase corresponds to
spin-singlet pairs of electrons (singlet bipolarons) residing on every other
lattice site.

We are not aware of any previous results for the phonon spectral function of
the 1D Holstein-Hubbard model.  The phonon spectrum of the 1D, half-filled
spinless Holstein model was studied in
Refs.~\cite{SyHuBeWeFe04,CrSaCa05,SyHuBe05,Hohenadler06,Sykora06}.  Numerical
results were obtained using exact diagonalization (restricted to small
cluster sizes) \cite{Hohenadler06}, cluster perturbation theory (with
artifacts related to the inherent translation symmetry breaking)
\cite{Hohenadler06}, and QMC (restricted in cluster size and temperature, and
with a Trotter error) \cite{CrSaCa05}. Regarding analytical work, the
projector-based renormalization group approach
\cite{SyHuBeWeFe04,SyHuBe05,Sykora06} provides the most comprehensive
results. For $\om_0<t$, the renormalization of the phonon spectrum at large
$q$ was studied by means of the dynamic charge structure factor
\cite{Hohenadler10a,PhysRevB87.075149}.  A general feature of Holstein models
is that for $\lambda=0$, the phonon dispersion $\om_q=\om_0$ is
trivial. However, for nonzero couplings, a renormalization takes place and
$\om_q\mapsto\tilde{\om}_q$. For the spinless Holstein model, $\tilde{\om}_q$
has been calculated in Refs.~\cite{SyHuBeWeFe04,Sykora06}.

\section{Method}\label{sec:method}

The CT-INT QMC method permits one to simulate rather general fermionic actions
(including retarded and nonlocal interactions) \cite{Rubtsov05}. Exploiting
the fact that the phonon degrees of freedom can be integrated out exactly to
obtain a fermionic action \cite{Assaad07,Feynman55}, it was applied to a
number of different electron-phonon problems
\cite{Assaad07,Assaad08,Hohenadler10a,Ho.As.Fe.12,PhysRevB87.075149,PhysRevB.88.064303,WeAsHo15_I}.
The method relies on the imaginary-time path-integral formulation of the
partition function, which is calculated using a weak-coupling perturbation
expansion \cite{Assaad07}. A strong-coupling formulation also exists, but is
less suited for lattice problems \cite{werner:146404}.

For the model~(\ref{eq:model-holsteinspinful}), the path-integral
representation of the partition function takes the form
\begin{align} \label{partitionfunction}
 Z = \int \mathcal{D}(\bar{c},c) \ e^{-S_0\left[\bar{c},c\right]-S_1\left[\bar{c},c\right]} \int \mathcal{D}(q) \ e^{-S_\text{ep}\left[\bar{c},c,q\right]} \,,
\end{align}
where we used the coherent-state representation $\hat{c}_{i\sigma} |c\rangle
= c_{i\sigma} |c\rangle$ with Grassmann variables $c_{i\sigma}$ for the
fermions, and the real-space representation $\hat{Q}_i |q\rangle = q_i
|q\rangle$ for the phonons. We split the action into 
the free-fermion part $S_0$, the Hubbard interaction $S_1$, 
and the remainder $S_\text{ep}$ containing the free-phonon part as well as
the coupling of the displacement fields to the electrons, given by
\begin{align}
  S_\text{ep}  \hspace*{-.2em} =  \hspace*{-.2em}\int_0^{\beta} \hspace*{-.25em}d\tau \sum_i
  \Big\{
    \mbox{$\frac{M}{2}$}
    \dot{q}_i^2(\tau)
   + \mbox{$\frac{K}{2}$}  q_i^2(\tau)
    - g q_i(\tau)  \rho_i(\tau) 
   \Big\}\,.
\end{align}
Integration over the fields $q$ in Eq.~(\ref{partitionfunction})
leads to an effective fermionic action $S_2$ that can be simulated with the CT-INT
method in the same way as fermionic models \cite{Assaad07}.

\begin{figure*}[t]
  \includegraphics[width=1\textwidth]{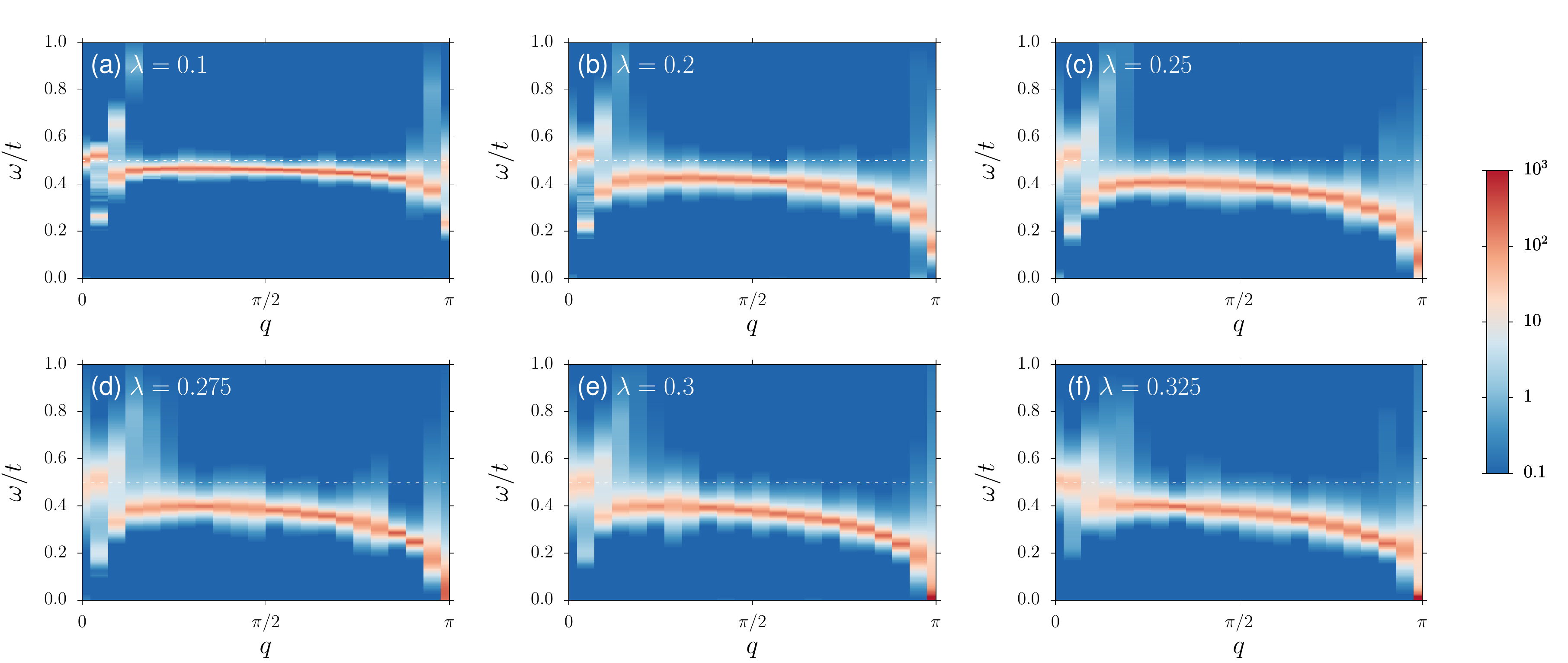}
  \caption{\label{fig:spinful-lambda} (Color online) Phonon spectral function
    $B(q,\omega)$ of
    the spinful Holstein model. Dashed lines correspond to $\om_0=0.5t$.  Here, $U=0$,
    and $\beta t=L=42$. The Peierls transition occurs at
    $\lambda_c\approx0.25$ \cite{hardikar:245103,0295-5075-84-5-57001}. Color scheme based on Ref.~\cite{colorscheme}.}
\end{figure*}

Previous applications of the CT-INT method were restricted to correlation
functions of fermionic operators. By using a generating functional with
source fields that couple to the lattice displacement fields, we can derive
an estimator for the phonon propagator $D_{ij}(\tau)$ in terms of the
time-displaced charge correlation function.  Explicitly, to measure the
phonon propagator
\begin{align}
 D_{ij}(\tau)=\langle q_i(\tau) q_j(0) \rangle \,,
\end{align}
we add a source term $S^{\eta} = - \int d\tau \sum_i \eta_i(\tau) q_i(\tau)$
to $S_\text{ep}$, where $\eta_i(\tau)$ is a real field.  After integrating
out the displacement fields, we arrive at an effective action describing the
electron-phonon interaction,
\begin{align}\label{effaction}
 S^\eta_2 &= - \frac{g^2}{2} \iint_{0}^{\beta} d\tau d\tau' \sum_i
  \left[\rho_i(\tau) + g^{-1} \eta_i(\tau) \right] \\\nonumber
 &\hspace*{7em}\times D^0(\tau-\tau')
  \left[\rho_i(\tau') + g^{-1}  \eta_i(\tau') \right]\,.
\end{align}
The appearance of the free phonon propagator
\begin{align}
 D^0(\tau)
   = \frac{1}{2 M \omega_0} \frac{ \cosh{\left[\omega_0 ( \beta/2 - |\tau| )\right]} }{ \sinh{\left[ \omega_0 \beta / 2 \right]}}\,,
\end{align}
defined for $-\beta \leq \tau \leq \beta$, leads to a retarded interaction in imaginary time.
The interacting propagator $D_{ij}(\tau)$ can be obtained from
Eq.~(\ref{effaction}) via a functional derivative with respect to
$\eta_i(\tau)$ in the limit $\eta\rightarrow0$. The result,
\begin{align}\label{phononpropint}
  D_{ij}(\tau) &=   
  D^0(\tau)\delta_{ij}
  + g^2 
   \iint_{0}^{\beta} d\tau_1 d\tau_2 
  D^0(\tau_1-\tau) \\ \nonumber
  &\hspace*{9em}\times  
  D^0(\tau_2) 
  \left\langle
    \rho_i(\tau_1) \rho_j(\tau_2)
  \right\rangle\,,
\end{align}
is the sum of the free propagator and an interaction term that involves the
time-displaced charge correlation function via a double convolution with
$D^0(\tau)$. The charge correlation function can be accurately measured with
the CT-INT method. Equation~(\ref{phononpropint}) has previously been derived
for the Anderson-Holstein impurity model \cite{0953-8984-14-3-312}.

We evaluated Eq.~(\ref{phononpropint}) for each bin average of $\left\langle
  n_{i}(\tau) n_{j}(0) \right\rangle$, and used a binning analysis to obtain
reliable statistical errors. To minimize systematic errors in the numerical
integration, we carried out one of the integrals in Eq.~(\ref{phononpropint})
analytically by exploiting the periodicity of $D^0(\tau)$ and $\left\langle
  n_{i}(\tau) n_{j}(0) \right\rangle$.  Measurements were made on an
equidistant grid on the imaginary time axis with $\Delta\tau t=0.1$.  From
$D_{ij}(\tau)$ we obtained the phonon spectral function (we define
$\Delta_{nm}=E_n-E_m$)
\begin{equation}
  B(q,\omega) = 
  \frac{1}{Z} \sum_{n,m} | \langle m | \hat{Q}_q | n\rangle |^2 e^{-\beta
  E_m}  
  \delta(\omega - \Delta_{nm}) 
\end{equation}
by carrying out a Fourier transformation and using the stochastic maximum
entropy method \cite{Beach04a} for the analytic continuation. The use of
$\rho_i(\tau)$ instead of $n_i(\tau)$ in Eq.~(\ref{phononpropint})
amounts to subtracting the $T=0$ static contribution to $B(q=0,\omega)$.
Our results obey the sum rule $\int_0^\infty d\omega B(q,\omega) (1 +
e^{-\beta \omega})= D(q,\tau=0)$. Additional sum rules are discussed in the
appendix.

\section{Numerical results}\label{sec:results}

We first consider the spinful Holstein model
[Eq.~(\ref{eq:model-holsteinspinful}) with $U=0$] with $\omega_0=0.5t$, which
exhibits a Peierls metal-insulator transition at $\lambda_{c}\approx0.25$
\cite{hardikar:245103,0295-5075-84-5-57001}.  Figure~\ref{fig:spinful-lambda}
shows the phonon spectral function $B(q,\omega)$ for different values of
$\lambda$. The system size and temperature were chosen as $L=\beta t=42$ (see
also Sec.~\ref{sec:discussion}).

For a weak coupling $\lambda=0.1$, Fig.~\ref{fig:spinful-lambda}(a) reveals
that the bare phonon mode at $\om_q=\om_0$ is renormalized near $q=0$ and
$q=\pi$, but largely unaffected at intermediate $q$. The small-$q$ feature
will be explained as a hybridization effect in Sec.~\ref{sec:discussion}.
For a stronger coupling $\lambda=0.2$, shown in
Fig.~\ref{fig:spinful-lambda}(b), the renormalization of the phonon mode is
significantly more pronounced, and we observe a partial softening near
$q=\pi=2\kF$, which is a precursor of the Peierls transition.  Upon
increasing $\lambda$ further, the phonon spectrum becomes gapless close to
$\lambda_c=0.25$. Beyond $\lambda_c$, the results in
Figs.~\ref{fig:spinful-lambda}(d)--\ref{fig:spinful-lambda}(f) are consistent (see below) with the
emergence of a central ($\om=0$) peak at $q=\pi$, and a second peak at
$\om>0$. The central peak carries substantial spectral weight (which diverges
with system size) and reflects the long-range lattice order with modulation
$q=2\kF=\pi$. The phonon dispersion throughout the Brillouin zone appears to
be most pronounced for $\lambda\approx\lambda_c$, and becomes flatter in the
Peierls phase.

\begin{figure}[t]
  \includegraphics[width=0.4\textwidth]{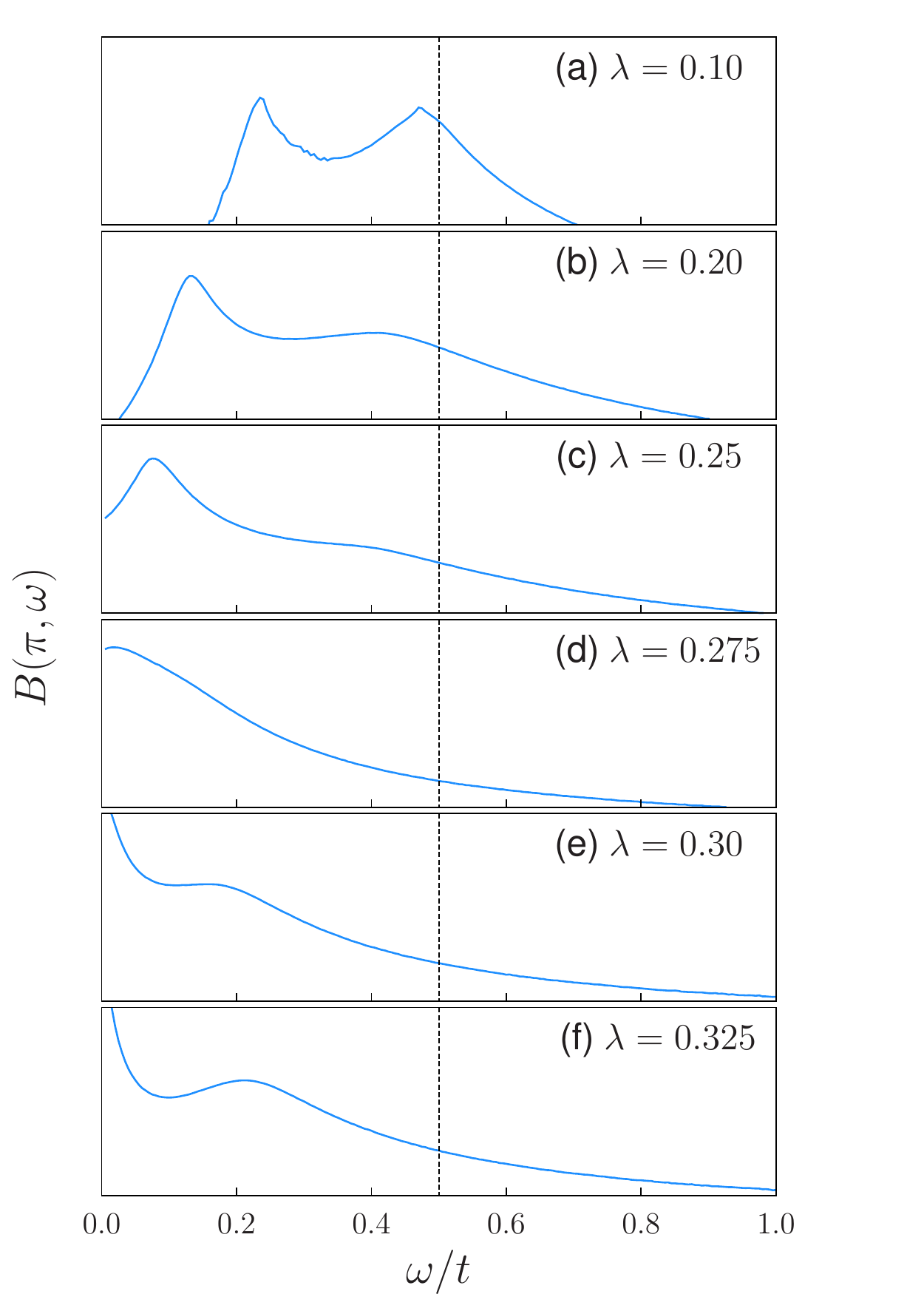}
  \caption{\label{fig:spinful-lambda-pi} (Color online) Phonon spectral
    function $B(q,\omega)$ at $q=\pi$ for the same parameters as in
    Fig.~\ref{fig:spinful-lambda}. The dashed vertical lines indicate
    $\omega_0=0.5t$. As in Fig.~\ref{fig:spinful-lambda}, we use a
    logarithmic scale and a plot range $[0.1,10^3]$.}
\end{figure}

To understand the evolution of $B(q,\omega)$ at the ordering wavevector
$q=\pi$ in more detail, we show in Fig.~\ref{fig:spinful-lambda-pi} the
spectrum $B(\pi,\omega)$ for the same parameters as in
Fig.~\ref{fig:spinful-lambda}. Except for $\lambda=0.275$, we observe two
separate low-energy peaks in the phonon spectrum, although they are difficult
to discern in the density plots of Fig.~\ref{fig:spinful-lambda}. Starting at
weak coupling $\lambda=0.1$, there is a peak at $\omega\approx 0.2t$
(corresponding to the finite-size charge gap in the dynamic charge structure
factor at $q=\pi$, which scales as $1/L$), and a peak close to the
bare phonon frequency $\omega=\omega_0=0.5t$. With increasing $\lambda$, both
peaks move toward smaller $\omega$. For the system size used, the spectrum
becomes gapless close to the critical coupling $\lambda_c\approx 0.25$,
coinciding with the appearance of the central
peak. Figures~\ref{fig:spinful-lambda-pi}(e)
and~\ref{fig:spinful-lambda-pi}(f) reveal that a second peak emerges that
becomes harder with increasing $\lambda$ but has small spectral weight
compared to the central peak.

\begin{figure}[t]
  \includegraphics[width=0.375\textwidth]{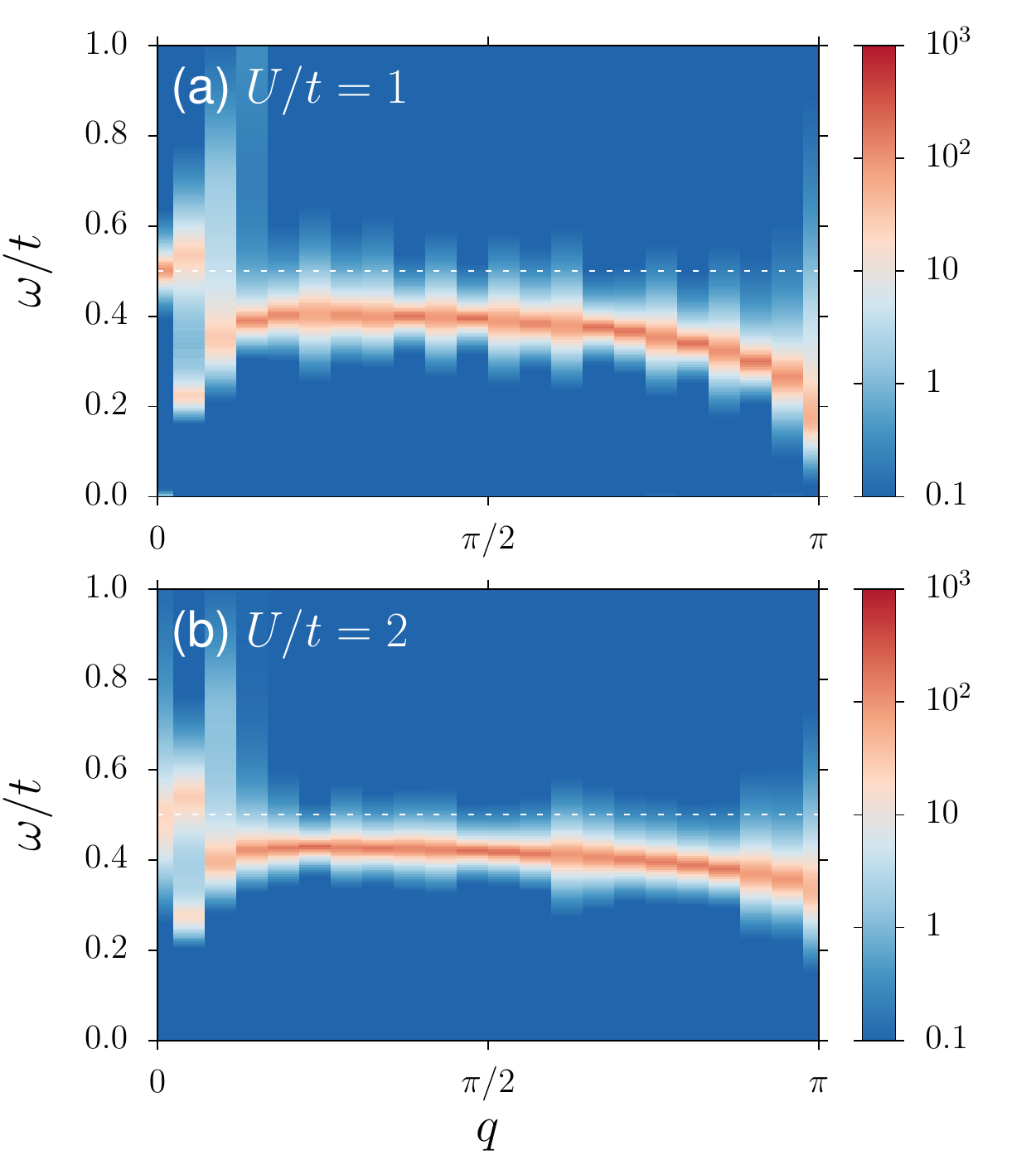}
  \caption{\label{fig:spinful-U} (Color online) Phonon spectral function
    $B(q,\omega)$ of
    the Holstein-Hubbard model. The dashed line corresponds to $\om_0=0.5t$.
    Here, $\lambda=0.3$, and $\beta t=L=42$.}
\end{figure}

It is also interesting to consider the impact of Coulomb repulsion in the
framework of the Holstein-Hubbard model.  In Fig.~\ref{fig:spinful-U}(a) we
show the phonon spectral function for $\lambda=0.3$ and $U=t$, which should
be compared to the results of Fig.~\ref{fig:spinful-lambda}(e). Quite
remarkably, we find a comparable renormalization of the bare phonon mode, but
no central peak. The absence of the latter is expected since $U=t$ is
sufficient to drive the system from the Peierls into the intermediate
metallic phase, cf. Fig.~8(a) in Ref.~\cite{hardikar:245103}. On the other
hand, the renormalization of the phonon mode is much stronger than expected
based on a simple effective Hubbard model with interaction
$U_\text{eff}=U-4t\lambda$, which is justified for sufficiently large
$\omega_0/t$. For the parameters of Fig.~\ref{fig:spinful-U}(a), we have
$U_\text{eff}=-0.2t$. However, the phonon spectrum is renormalized more
strongly than in Fig.~\ref{fig:spinful-lambda}(a), where $\lambda=0.1$ and
$U_\text{eff}=-0.4t$, suggesting that not only the effective interaction but
also the bare electron-phonon coupling determines the phonon renormalization.
Finally, for $U=2t$ [Fig.~\ref{fig:spinful-U}(b)], corresponding to the Mott phase
\cite{hardikar:245103}, the renormalization of the phonon mode is
significantly weaker, and the phonon spectrum resembles quite closely the
weak-coupling results shown in Fig.~\ref{fig:spinful-lambda}(a).

\section{Discussion}\label{sec:discussion}

In this section, we relate our numerical results to previous work and to
theoretical expectations. Additionally, we provide an explanation of the
small-$q$ feature observed in the metallic phase, and show how the discrete
symmetry spontaneously broken at the Peierls transition can be captured in a
continuum field theory.

\subsection{Relation to the charge structure factor}

As illustrated by the exact relation~(\ref{phononpropint}), the phonon Green
function and the time-displaced charge correlation function are related by a
convolution. Consequently, the phonon spectral function $B(q,\omega)$ and the
dynamic charge structure factor
\begin{align}\label{eq:nqw}
  N(q,\om)
  &=
  \frac{1}{Z}\sum_{n,m} {|\langle {m}| \hat{\rho}_q |{n}\rangle|}^2
  e^{-\beta E_m} 
  \delta(\om-\Delta_{nm})\,, 
\end{align}
with $\hat{\rho}_q= L^{-1/2}\sum_r e^{iqr} \hat{\rho}_r$, contain in
principle the same information, although the spectral weights may differ by
orders of magnitude. In particular, as discussed in more detail in the Appendix,
the dynamic charge structure factor also
reveals the renormalized phonon excitations
\cite{Assaad08,Hohenadler10a,PhysRevB87.075149}, and the phonon spectral
function also contains signatures of the particle-hole continuum.

Importantly, for values of $q$ where the bare phonon mode and the
particle-hole continuum overlap, the renormalized phonon mode
$\tilde{\omega}_q$ will be damped, in contrast to the $\delta$-function
contribution suggested by the approximation for $B(q,\omega)$ given in
Eq.~(5) of Ref.~\cite{Sykora06}. For intermediate phonon frequencies
$\omega_0\approx t$, no clear separation exists between the phonon mode and
particle-hole excitations. Finally, for $\omega_0\gg t$ (that is, larger than
the bare bandwidth $4t$ of the particle-hole continuum), or deep in the Peierls
phase where the lower edge of the particle-hole continuum lies above 
$\tilde{\omega}_q$, we expect the phonon
mode to remain a separate and well-defined excitation near $q=2\kF$, as
suggested by Figs.~3(b) and~2(a) in Ref.~\cite{Sykora06}, respectively.

An important corollary of the relation between $B(q,\omega)$ and
$N(q,\omega)$ concerns the shape of the spectrum in the vicinity of
$q=\pi$. In the Peierls phase, the size of the unit cell (Brillouin zone)
doubles (halves). Although a perfect symmetry (extending to the spectral
weights of excitations) between $q=0$ and $q=\pi$ is only achieved in the
limit $\lambda\to\infty$, this doubling implies that the renormalized phonon
frequency $\tilde{\om}_q$ cannot extrapolate to zero at $q=\pi$ for
$\lambda>\lambda_c$. Such a gapless mode would necessarily have a counterpart
near $q=0$, and also in $N(q,\om)$.  However, gapless, long-wavelength
particle-hole excitations are not compatible with an insulating Peierls
state. We will see below that a simple low-energy theory instead suggests a
gapped renormalized phonon mode and an isolated central peak at $q=\pi$. This
picture is consistent with our numerical results. On the other hand, we
attribute the apparent existence of a gapless mode near $q=\pi$ in results
for the spinless Holstein model \cite{Hohenadler10a} to a failure to resolve
the expected two-peak structure in the dynamic charge structure factor.

\subsection{Origin of the small-$q$ feature}

\begin{figure}[t]
  \includegraphics[width=0.375\textwidth]{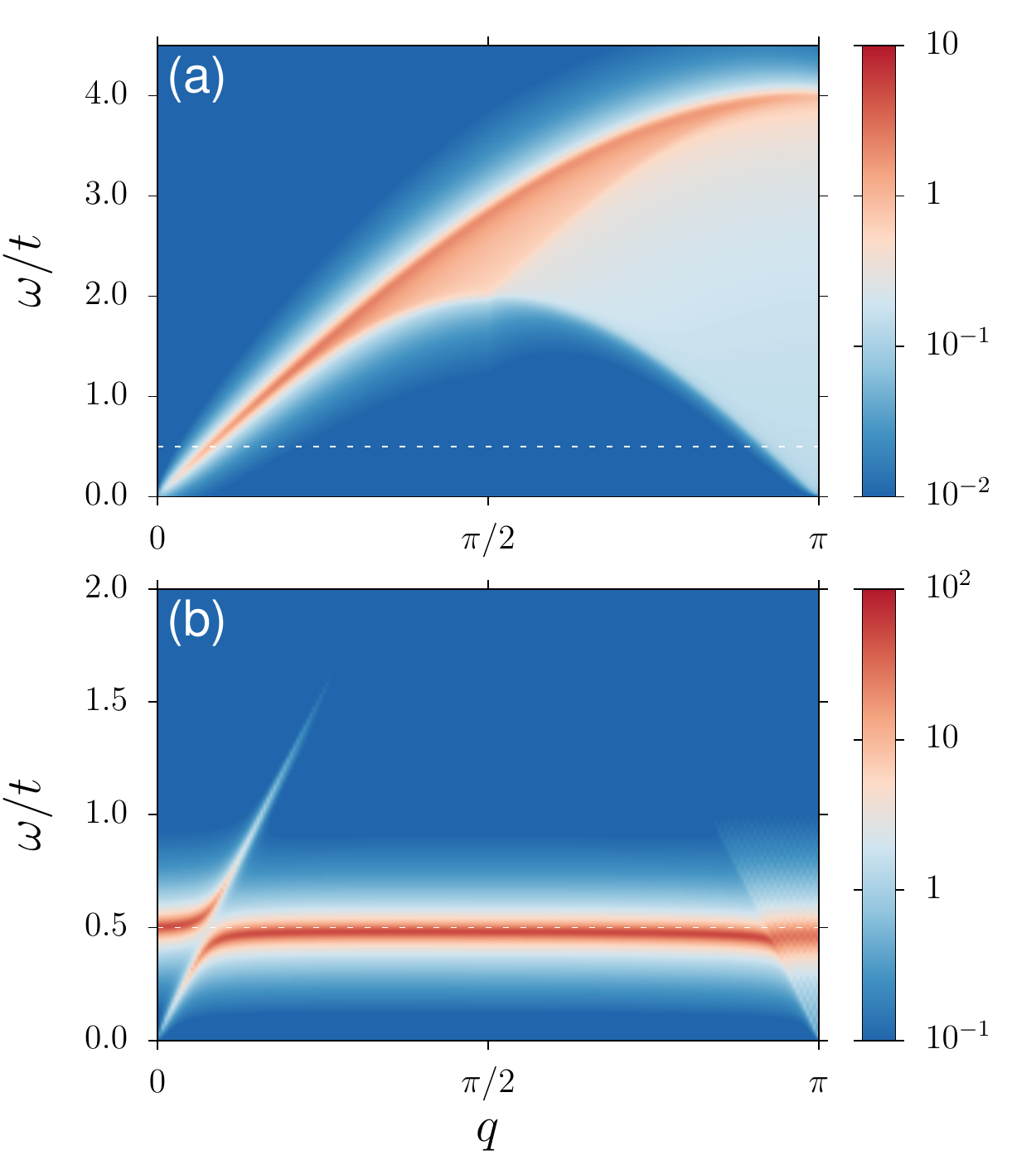}
  \caption{\label{fig:rpa} (Color online) (a) Dynamic charge
    structure factor of the noninteracting system ($\lambda=U=0$). (b) Phonon spectral
    function from Eq.~(\ref{eq:rpa}) for $\lambda=0.2$. The dashed lines indicate
    $\om_0=0.5t$. Here, $T=0$, and $L=402$.
  }
\end{figure}

The coupling between the bare phonon mode and the particle-hole excitations
of $N(q,\omega)$ [see Fig.~\ref{fig:rpa}(a)] via the electron-phonon
interaction provides an explanation of the small-$q$ feature visible in
Fig.~\ref{fig:spinful-lambda}(a), as well as in the renormalized phonon
frequency $\tilde{\om}_q$ in Refs.~\cite{SyHuBeWeFe04,SyHuBe05}. In the
analytical results of Refs.~\cite{SyHuBeWeFe04,SyHuBe05}, this feature occurs
at a nonzero $q$ and is quite sharp in momentum space, making it difficult to
resolve fully in our finite-size data and explaining its absence in previous
exact diagonalization results for small clusters \cite{Hohenadler06}.

Here, we explain this feature in terms of the hybridization between the bare
phonon frequency $\om_q=\om_0$ and particle-hole excitations that---in one
dimension---have the linear dispersion $\om= v_\text{F} q$ for small $q$, see
Fig.~\ref{fig:rpa}(a). The hybridization is captured by the random phase
approximation for the phonon propagator,
\begin{equation}\label{eq:rpa}
  D^{-1}(q,i\Omega_n) = [D^0(i\Omega_n)]^{-1} + \lambda^2 \chi^0(q,i\Omega_n)\,,
\end{equation}
where $\Omega_n$ is a bosonic Matsubara frequency, and $\chi^0(q,i\Omega_n)$
is the noninteracting charge susceptibility. Taking $\lambda$ as a free
parameter to match the numerical results, we find a hybridization at the
intersection point of the free phonon dispersion and the free particle-hole
excitations, as well as a gapless linear mode below the bare dispersion that
corresponds to long-wavelength charge fluctuations. While the CT-INT results
in, for example, Fig.~\ref{fig:spinful-lambda}(a), do not fully resolve the
hybridization, the agreement is satisfactory. Note that the random phase
approximation does not take into account any phonon softening related to
charge order, or the renormalization of ${v}_\text{F}$ by interactions. 
The hybridization of charge and phonon modes at small $q$ also
follows from a Tomonaga-Luttinger model \cite{PhysRev.136.A1582}.

In accordance with the analytical results of Refs.~\cite{SyHuBeWeFe04,SyHuBe05},
a hybridization of the phonon mode with the particle-hole continuum is also
visible near $q=\pi$, and gives rise to damping of the phonon excitations.
The above explanation suggests that the small-$q$ feature is absent deep in
the insulating Peierls phase, because the latter does not have low-energy
excitations near $q=0$. Accordingly, a suppression of the hybridization
feature with increasing $\lambda$ is visible in Fig.~\ref{fig:spinful-lambda}.
Finally, within the random phase approximation, the hybridization does not
explain the observed hardening of the phonon mode near the zone boundary 
for $\om_0= 4t$ \cite{Hohenadler06}.

\subsection{Finite-size effects}

Although our system size of $L=42$ is significantly larger than in previous
work, it is important to separate finite-size effects from generic features.

The most notable finite-size artifact in our results is the charge gap at
$q=\pi$ in the dynamic charge structure factor and (for $\lambda>0$) also in
the phonon spectral function. In the noninteracting case, this gap scales as
$1/L$. While it is negligibly small in Fig.~\ref{fig:rpa}(a) ($L=402$), it is
about $0.3t$ for the system size $L=42$ used in our simulations. The results
for the phonon spectral function in Fig.~\ref{fig:spinful-lambda-pi} reveal
that with increasing $\lambda$, the charge gap is reduced. At the critical
point, the central peak at $\omega=0$ appears. The closing of the finite-size
charge gap almost exactly at the critical coupling $\lambda_c$ is not
expected to be generic, but to depend on system size and parameters.

While a charge gap is expected in $B(\pi,\omega)$ in the Peierls phase even
in the thermodynamic limit, both $B(q,\omega)$ and $N(q,\omega)$ are gapless
at $q=\pi$ throughout the metallic phase for $L\to\infty$ \cite{Sykora06}. In the insulating
Peierls phase, the central peak at $\omega=0$ is separated from excitations
with $\om>0$ by the interaction-generated charge gap. Importantly, the lower edge of the
particle-hole continuum in the metallic phase corresponds to a branch cut
(similar to other excitations of 1D systems) with finite
spectral weight, whereas the weight of the central peak at $\omega=0$ in the
Peierls phase diverges in the thermodynamic limit. Finite-size effects on the
phonon spectrum of a spin-Peierls model have been discussed in
Ref.~\cite{Michel07}.

To avoid spurious finite-size effects, the system size should be large enough
to have a noninteracting charge gap smaller than the bare phonon frequency. Otherwise, there
is no coupling between the bare mode and the particle-hole continuum, which
is not generic for the adiabatic regime. In the latter, and more generally
for any $\omega_0<4t$, the bare phonon mode lies on top of the noninteracting
particle-hole continuum in extended regions of the Brillouin zone. A nonzero
electron-phonon coupling then gives rise to a mixing of these excitations,
leading to renormalization and a finite lifetime of phonon excitations. In
this sense, Eq.~(5) in Ref.~\cite{Sykora06} should be regarded as an
approximate result because it suggests the existence of phonon excitations
with infinite lifetime described by $\delta(\om-\tilde{\om}_{q})$.

\subsection{Soft-mode versus central-peak behavior}

Peierls transitions are often classified as either soft-mode or central-peak
transitions. In terms of the coupling $\lambda$ (rather than temperature, as
appropriate for experiments on quasi-1D systems), a 1D soft-mode transition
involving a single phonon mode is characterized by a softening
$\omega_{2\kF}\to0$ for $\lambda<\lambda_c$, a completely soft mode
$\omega_{2\kF}=0$ at $\lambda=\lambda_c$, and a subsequent hardening (\ie,
$\omega_{2\kF}>0$) in the Peierls phase with an additional central peak at
$\omega=0$ reflecting the long-range lattice order. In contrast, for a
central-peak transition, $\omega_{2\kF}$ stays nonzero or even hardens across
the transition, and a central peak appears at the critical point and persists
for $\lambda>\lambda_c$.

The above simple picture is modified in several ways. (i) Because of the
relation to the charge structure factor, $B(q,\omega)$ in general has a
continuum of excitations. Although the spectra shown here for the adiabatic
regime are dominated by a few peaks, the particle-hole continuum has
substantial spectral weight for $\omega_0>t$ \cite{Hohenadler06,Sykora06}.
(ii) In the thermodynamic limit, $B(q,\omega)$ has soft (\ie, gapless)
excitations in the metallic phase corresponding to the gapless particle-hole
continuum at $q=\pi$. (iii) In the Peierls phase, the renormalized phonon
mode (which has a finite energy for $\lambda>\lambda_c$) has very small
spectral weight compared to the divergent central peak, which may make its
identification difficult for both numerical simulations and
experiments. Similar behavior has been reported for a spin-Peierls model in
Ref.~\cite{Michel07}; in particular, it was pointed out that the two peaks in
$B(q,\pi)$ are only visible on large enough system sizes. Finally, it is not
clear if the hardening of $\om_{2\kF}$ in the Peierls phase can be clearly
distinguished from the opening of the gap in the particle-hole continuum
in the vicinity of $\lambda_c$.

Keeping in mind these complications, our numerical results in
Figs.~\ref{fig:spinful-lambda} and~\ref{fig:spinful-lambda-pi} are
nevertheless compatible with a soft mode transition. The dispersive feature
that exists throughout the Brillouin zone and emerges from the bare phonon
mode $\omega_q=\omega_0$ upon turning on the electron-phonon interaction can
be identified with the renormalized phonon frequency $\tilde{\omega}_q$. We
observe that $\tilde{\omega}_q$ softens near $q=\pi$ as
$\lambda\to\lambda_c$, and hardens again in the Peierls phase. However, due
to limitations in system size and spectral resolution, we are unable to
unambiguously demonstrate the existence of a true soft mode at $\lambda_c$.

Let us briefly discuss relevant previous works on this issue. 
In Refs.~\cite{SyHuBeWeFe04,Sykora06,SyHuBe05}, the renormalized phonon
frequency $\tilde{\omega}_{q}$ of the spinless Holstein model
was found to soften at the critical point, and harden
again in the Peierls phase. This behavior is consistent with a soft-mode
transition, but the results for the phonon spectral function in Fig.~2 of
Ref.~\cite{Sykora06} do not seem to show a central peak related to long-range
order, in contrast to numerical results \cite{Hohenadler06}. A softening is
also visible in exact diagonalization results for small clusters in
Ref.~\cite{Hohenadler06}. Based on QMC simulations and fits of the phonon
propagator to a simple two-mode form, a softening (hardening) in the metallic
(insulating) phase was suggested in Ref.~\cite{CrSaCa05}. Although we can
reproduce these results, it is not clear how reliable such fits are given the
large ratio of spectral weights between the central peak and the phonon peak.
Moreover, we find that the minimum of the fitted phonon frequency does not
give the correct critical value $\lambda_c$ in the adiabatic regime
($\om_0=0.1t$).  Finally, the phonon softening and ensuing hardening observed
in Migdal-Eliashberg theory \cite{Migdal58} and in approximate solutions of
the Holstein model \cite{AlKaRa94} for any band filling appear to be a
consequence of the breakdown of the adiabatic approximation. A soft
phonon mode seems to be always linked with a lattice instability (phase
transition), which requires a commensurate band filling.

\subsection{Field-theory description}

The observation of a Peierls transition implies the existence of a $Z_2$
symmetry which, in one dimension, can be spontaneously broken at zero
temperature and would account for the observed central peak.  Here we provide
a minimal continuum theory that reveals such a $Z_2$ symmetry for the
Holstein model.  For simplicity, we consider spinless fermions and assume an
arbitrary value of the Fermi wave vector $k_\text{F}$.

Near $\pm\kF$, the electronic degrees of freedom can be described
by a linear dispersion, leading to the Dirac form
\begin{equation}\label{eq:app1}
  \hat{H}_\text{e}    =  v_\text{F} \sum_{k}    k {\Psi}^{\dagger}_{k} {\sigma}_z   {\Psi}^\nag_{k}\,,
\end{equation}
where ${\Psi}^{\dagger}_{k} = (\hat{L}^{\dagger}_{k}, \hat{R}^{\dagger}_{k})
$ is a spinor of creation operators for left and right moving fermions,
respectively, and $\sigma_z$ is the usual Pauli matrix. $\hat{H}_\text{e}$
has a $U_{S_z}(1)$ symmetry related to the pseudospin $\hat{S}_z =
\frac{1}{2}\sum_{k} \Psi^{\dagger}_{k} \sigma_z \Psi^\nag_k$; we have
$\hat{U}^{-1}_{\theta} \Psi_{k} \hat{U}_{\theta} = {U}_{\theta} {\Psi}_{k}$
with $ \hat{U}_{\theta} = e^{i \theta \hat{S}_z} $ and $ {U}_\theta = e^{i
  (\theta/{2}) {\sigma}_z } $.

The free phonon part can be written as
\begin{equation}
  \hat{H}_\text{p}   =     \sum_{k}       \omega^\nag_{k}
  \hat{\alpha}^{\dagger}_{k} \hat{\alpha}^\nag_{k} 
\end{equation}
with $\omega_k=\om_0$ for optical phonons.
 
In the framework of bosonization, the Peierls transition arises from
phonon-mediated single-particle backscattering described by (see Appendix~B
of Ref.~\cite{Assaad08})
\begin{eqnarray}\label{eq:app2}
  \hat{H}_\text{I} = \frac{\tilde{g}}{\sqrt{L} } \sum_{k,q}   & &  \left[  \hat{R}^{\dagger}_{k}\hat{L}^\nag_{k-q}   \left(   \hat{\alpha}^{\dagger} _{-2\kF -q}  +  \hat{\alpha}^\nag_{2\kF + q} \right)  \right. \\ \nonumber
  & &  + \left.  \hat{L}_{k-q}^{\dagger} \hat{R}^\nag_{k}   \left( \hat{\alpha}^{\dagger}_{2\kF + q}   +   \hat{\alpha}^\nag_{-2\kF-q}  \right) \right]\,.   
\end{eqnarray}

To derive a low-energy theory, we only consider small momenta $|q|\ll 2\kF$.
Consequently, for our purposes, $\hat{\alpha}^{\dagger} _{-2\kF -q}$ and
$\hat{\alpha}^\nag_{2\kF + q} $ may be regarded as independent bosonic modes
\footnote{Our field theory implies a continuum limit such that
  momentum conservation is exact. Hence the identification of $2\kF$ to
  $-2\kF$ for the half-filled band is not justified.}.  With
${\alpha}^{\dagger}_{q} = ( \hat{\alpha}^{\dagger}_{-2\kF -q},
\hat{\alpha}^{\dagger}_{2\kF +q})$ the phonon Hamiltonian can be written as
\begin{equation}
  \hat{H}_\text{p} = \sum_{q}    {\alpha}^{\dagger}_{q}     \left[ \omega_{+}(q) +\omega_{-}(q) {\sigma}_z  \right]  {\alpha}^\nag_{q} \,,
\end{equation}
where $ \omega_{\pm}(q) = \left( \omega_{-2\kF -q} \pm \omega_{2\kF +q}
\right)/2$. Under the above assumption of two independent modes,
$\hat{H}_\text{p}$ has two $U(1) $ symmetries generated by the conservation
of the total phonon number $ \hat{N} = \sum_{q} {\alpha}^{\dagger}_{q}
{\alpha}^\nag_{q} $, and of $ \hat{N}_z = \sum_{q} {\alpha}^{\dagger}_{q}
{\sigma}_z {\alpha}^\nag_{q} $, denoted as $U_N(1)$ and $U_{N_z}(1)$,
respectively.

The electron-phonon interaction term~(\ref{eq:app2}) breaks the full $U_N(1)
\times U_{N_z}(1) \times U_{S_z}$(1) of the free Hamiltonian (in particular,
the number of phonons is not conserved). However, under a combined
transformation $\hat{U}_{\phi,\theta}= e^{i \phi \hat{N}_z} e^{i
  \theta\hat{S}_z} $ the interaction term transforms as
\begin{eqnarray}
  \hat{U}^{-1}_{\phi,\theta}  \hat{H}_\text{I} \hat{U}^\nag_{\phi,\theta}   =  
  \frac{\tilde{g}}{\sqrt{L} } \sum_{k,q}    && \Psi^{\dagger}_{k}
  {U}^{\dagger}_{\theta} {\sigma}_{-} {U}_{\theta}^\nag
  \Psi_{k-q}^\nag  e^{i \phi} 
  \\\nonumber
  &&\times\left(   \hat{\alpha}^{\dagger} _{-2\kF -q}  +  \hat{\alpha}_{2\kF + q}  \right)   + \text{H.c.}\,,
\end{eqnarray} 
where $\sigma_{\pm} = \frac{1}{2} \left( \sigma_x \pm i \sigma_y
\right)$. Since $U^{\dagger}_{\theta} {\sigma}_{-} {U}^\nag_{\theta} = e^{i
  \theta} {\sigma}_{-} $, setting $ \theta = -\phi$ shows that ${\hat{S}_z -
  \hat{N}_z}$ is a generator of a $U_{S_z - N_z }(1)$ symmetry of the full
Hamiltonian $\hat{H}_\text{e}+\hat{H}_\text{p}+\hat{H}_\text{I}$. On the
other hand the interaction term reduces the $U_N(1)$ symmetry to a discrete
$Z_2$ symmetry.  Therefore, the symmetry of the full Hamiltonian is $U_{S_z -
  N_z }(1)\times Z_2$.

As mentioned previously, the $U_{S_z - N_z }(1) \times Z_2$ symmetry can
explain the existence of a Peierls state with long-range order at $T=0$.  The
$U_{S_z - N_z }(1) $ symmetry accounts for a combined charge and phonon mode,
whereas the $Z_2$ symmetry, if broken at $T=0$, gives rise to the central
peak feature.  Within our continuum field-theory description, commensurate
band fillings cannot be unambiguously defined. Nevertheless, the reduction of
the $U_N(1)$ symmetry to $Z_2$ accounts for the expected pinning of the
charge-density wave for commensurate band fillings \cite{Lee74}.  The $U_{S_z
  - N_z }(1)$ symmetry is reminiscent of the $U(1)$ Gross-Neveu theory with
one bosonic mode and two fermion flavors \cite{PhysRevB.87.041401} that
describes, for example, the transition from a Dirac semimetal to a Kekule
ordered state in two dimensions \cite{PhysRevLett.111.066401}.

\section{Conclusions}\label{sec:conclusions}

We showed that the phonon spectral function of electron-phonon models can be
calculated using the CT-INT QMC method with the help of a generating
functional. The same idea can be used to measure other correlation functions
and observables, and for other models, which further extends the usefulness
and versatility of the CT-INT method.

Our results for the 1D Holstein-Hubbard model are consistent with a soft-mode
Peierls transition in the adiabatic regime, characterized by a softening of
the $q=2\kF$ phonon excitations, a gapless mode at the critical point, and a
subsequent hardening of the phonon mode in the Peierls phase.
However, this simple picture is complicated by finite-size effects, and by
the mixing of phonon and charge excitations mediated by the electron-phonon
interaction. We explained a small-$q$ anomaly of the metallic phase,
previously observed for the spinless Holstein model, in terms of a
hybridization of the bare phonon mode and the particle-hole continuum.
Overall, our results for the spinful Holstein model are very similar to
previous results for the spinless Holstein model.

For a Hubbard repulsion large enough to yield a metallic state,
we observed a suppression of the central peak related to long-range order,
but a significant remaining renormalization of the phonon mode that only
disappears in the Mott phase for even larger Hubbard interaction.

Finally, we provided a unified picture of current and previous work, and
showed how the discrete symmetry spontaneously broken in the Peierls phase
can be captured in a continuum field theory.

\begin{acknowledgments}
  We acknowledge access to the J\"ulich Supercomputing Centre, and
  financial support from the DFG Grants No. AS120/10-1 and No. Ho 4489/3-1 (FOR
  1807). We further thank J. Bhaseen, F. Goth, I. Herbut,
  J. Hofmann, F. Parisen Toldin, and G. Wellein for helpful discussions.
\end{acknowledgments}

\appendix*

\section{Exact relation between phonon and charge spectra}\label{appendix}

For the Holstein model, the phonon spectral function $B(q,\omega)$ and the
dynamic charge structure factor $N(q,\omega)$ in principle contain the same
information. Here, starting from Eq.~(\ref{phononpropint}), we derive an
exact relation between the spectral functions as well as additional sum
rules, and discuss the implications.

\subsection{Analytic properties of the spectral functions}

To simplify the notation, we define $\bar{B}(q,\omega) = (1-e^{-\beta
  \omega}) B(q,\omega)$ and $\bar{N}(q,\omega) = (1-e^{-\beta \omega})
N(q,\omega)$.  The phonon spectral function $\bar{B}(q,\omega)$ can be
obtained from the Lehmann representation of the phonon propagator
\begin{align}\label{phononLeh}
  D(q,z) = - \int_{-\infty}^{\infty} d\omega \ \frac{\bar{B}(q,\omega)}{z -
    \omega}
\end{align}
by analyzing its pole structure in the complex-frequency plane.
For simplicity, we restrict our considerations to finite Hilbert spaces,
where $D(q,z)$ has only simple poles on the real axis \footnote{We consider
  finite lattices with an arbitrary cutoff for the phonons.  The same results
  can be obtained by evaluating
  $\bar{B}(q,\omega)=\operatorname{Im}[D(q,\omega + i\eta)]/\pi$ for the
  general case, where $D(q,z)$ has a branch cut on the real axis.}
determined by the exact relation
\begin{align}\label{phononpropmom}
  D(q,z) = D^0(z) + g^2 D^0(z)^2 \chi(q,z) \, .
\end{align}
{The term $\sim g^2$ in Eq.~(\ref{phononpropmom}) 
gives rise to a product of poles arising from the free phonon propagator
$D^0(z)=M^{-1} (\omega_0^2 - z^2)^{-1}$ and the charge susceptibility
\begin{align}
  \chi(q,z) = \left\langle \rho_q(z)\rho_{-q}\right\rangle = -
  \int_{-\infty}^{\infty} d\omega \ \frac{\bar{N}(q,\omega)}{z - \omega} \,.
\end{align}
A partial-fraction decomposition and comparison of the pole
structure of the two sides of Eq.~(\ref{phononpropmom}) gives
\begin{align}\label{specrel}
  \bar{B}(q,\omega) = \bar{B}^0(q,\omega) + g^2 D^0(\omega)^2
  \bar{N}(q,\omega) + \bar{B}^1(q,\omega) \,.
\end{align}
For $g=0$, the phonon spectral function is given by 
\begin{align}
  \bar{B}^0(q,\omega) = \frac{1}{2M\omega_0} \left[ \delta(\omega-\omega_0) -
    \delta(\omega+\omega_0)\right] \,,
\end{align}
which describes excitations at the bare phonon frequency $\omega = \pm
\omega_0$.  Any finite electron-phonon coupling leads to the appearance of
two additional terms in Eq.~(\ref{specrel}): The first contains the whole charge spectrum
$\bar{N}(q,\omega)$ reweighted by the free phonon propagator, while
\begin{align}\label{B1}
  \begin{split}
    \bar{B}^1(q,\omega) = -&\frac{g^2}{\omega_0}  \left[ \delta(\omega-\omega_0) - \delta(\omega+\omega_0)\right] \\
    & \times \mathcal{P} \int_0^{\infty} d\omega' \omega' D^0(\omega')^2
    \bar{N}(q,\omega')
  \end{split}
\end{align}
gives an additional contribution at $\omega=\pm\omega_0$.  Here,
$\mathcal{P}$ denotes the principal value.

To derive Eq.~(\ref{specrel}), we used a partial-fraction decomposition,
leading to poles of both first and second order.  However, poles of second
order are forbidden by the Lehmann representation~(\ref{phononLeh}).
Therefore, their weights have to vanish, which (for $\omega_0>0$) leads to the sum rule 
\begin{align}\label{sumden}
  \mathcal{P} \int_{0}^{\infty} d\omega \
  \frac{\omega}{\omega_{\vphantom{0}}^2-\omega_0^2} \ \bar{N}(q,\omega) = 0
  \,.
\end{align}
From Eqs.~(\ref{specrel}) and~(\ref{sumden}), we obtain an equivalent sum
rule for $\bar{B}(q,\omega)$,
\begin{align}\label{sumph}
  \int_{0}^{\infty} d\omega \ \omega
  \left(\omega_{\vphantom{0}}^2-\omega_0^2\right) \bar{B}(q,\omega) = 0 \,,
\end{align}
which is just a combination of the first and third moment of
$\bar{B}(q,\omega)$.  In the same way, the absence of higher-order poles
requires $\bar{N}(q,\omega=\pm\omega_0)=0$.

For finite electron-phonon coupling, Eq.~(\ref{phononpropmom}) can also be
used to obtain the charge spectrum
\begin{align}\label{specrelrev}
  \bar{N}(q,\omega) = \frac{M^2}{g^2} \left( \omega_{\vphantom{0}}^2 -
    \omega_0^2\right)^2 \bar{B}(q,\omega)
\end{align}
from the phonon spectral function.  Here, contributions at
$\omega=\pm\omega_0$ are removed from $\bar{N}(q,\omega)$ by the prefactor.

\subsection{Implications for the spectral properties}

According to Eqs.~(\ref{specrel}) and~(\ref{specrelrev}), ${B}(q,\omega)$ and
${N}(q,\omega)$ share the same spectral information, up to an additional
contribution to $B(q,\pm\omega_0)$ that consists of the free phonon spectrum
$B^0(q,\omega)$ and a compensating term $B^1(q,\omega)$ due to finite
interactions.

For $q=0$, because of charge conservation, 
$N(q,\omega)$ only has a static contribution at $\omega=0$. Thus, $B^1(q=0,\omega)=0$ and the full phonon spectrum
is given by the free part at $\omega=\pm\omega_0$ and the static contribution
to $N(q,\omega)$.

For $q\neq0$, any finite electron-phonon coupling seems to shift the phonon
dispersion away from $\omega=\pm\omega_0$.  Exact diagonalization data for
the spinless Holstein model \cite{Hohenadler06} suggest that
$B(q,\pm\omega_0)$ vanishes and therefore $B^0(q,\omega)$ and $B^1(q,\omega)$
compensate each other \footnote{In general, this need not be the case.}.
In general, for $q\neq0$, both $B(q,\omega)$ and $N(q,\omega)$ contain signatures
of the phonon dispersion as well as the particle-hole continuum, although the
spectral weights may be very different.

The condition $B(q,\omega_0) \ge 0$ sets an upper bound to the integral in
Eq.~(\ref{B1}),
\begin{align}\label{denbound}
  \mathcal{P} \int_0^{\infty} d\omega
  \frac{\omega}{\left(\omega_{\vphantom{0}}^2-\omega_0^2\right)^2}
  \bar{N}(q,\omega) \le \frac{M}{2g^2} \,.
\end{align}
For the integral to converge, $N(q,\omega)$ has to vanish when approaching
$\omega_0$.  Thus, a nonzero electron-phonon interaction splits the charge
spectrum at $\omega=\omega_0$.

Further insight into the distribution of spectral weight can be obtained from
the sum rules~(\ref{sumden}) and~(\ref{sumph}).  We restrict our discussion
to $B(q,\omega)$, but the same arguments hold for $N(q,\omega)$.  For
$\omega>0$, $B(q,\omega)\geq 0$ but the prefactor $(\omega^2-\omega_0^2)$
changes sign at $\omega_0$.  This sign change divides the frequency axis into
regions $\omega<\omega_0$ and $\omega>\omega_0$, whose integrated spectral
weights have to compensate each other in the sum rule  \footnote{However,
the integrated weight $\int B(q,\omega) d\omega$ may be very different for $\omega<\omega_0$ and
  $\omega>\omega_0$.}. Note that spectral weight at $\omega=0$ and
$\omega=\omega_0$ does not contribute to the sum rule, therefore the
noninteracting phonon dispersion fulfills Eq.~(\ref{sumph}) trivially.  By
adiabatically switching on the electron-phonon coupling, the particle-hole
continuum enters $B(q,\omega)$ and spectral weight has to be redistributed to
fulfill Eq.~(\ref{sumph}).  For wave vectors such that the particle-hole
continuum only enters one of the two regions, spectral weight has to appear
in the other region.  This can be most easily achieved by shifting the phonon
dispersion.  Both the hardening of the phonon dispersion for $\omega_0 \gg
t$ \cite{Hohenadler06,Sykora06}, and the hybridization with
the particle-hole continuum as well as the phonon softening for $\omega_0 \ll
t$, are consistent with the sum rule~(\ref{sumph}).  Furthermore, in the Peierls phase, the
charge gap (the lowest excitation at $q=\pi$) cannot become larger than
$\omega_0$, as the central peak does not contribute to the sum rule~(\ref{sumph}).


\begin{thebibliography}{56}%
\makeatletter
\providecommand \@ifxundefined [1]{%
 \@ifx{#1\undefined}
}%
\providecommand \@ifnum [1]{%
 \ifnum #1\expandafter \@firstoftwo
 \else \expandafter \@secondoftwo
 \fi
}%
\providecommand \@ifx [1]{%
 \ifx #1\expandafter \@firstoftwo
 \else \expandafter \@secondoftwo
 \fi
}%
\providecommand \natexlab [1]{#1}%
\providecommand \enquote  [1]{``#1''}%
\providecommand \bibnamefont  [1]{#1}%
\providecommand \bibfnamefont [1]{#1}%
\providecommand \citenamefont [1]{#1}%
\providecommand \href@noop [0]{\@secondoftwo}%
\providecommand \href [0]{\begingroup \@sanitize@url \@href}%
\providecommand \@href[1]{\@@startlink{#1}\@@href}%
\providecommand \@@href[1]{\endgroup#1\@@endlink}%
\providecommand \@sanitize@url [0]{\catcode `\\12\catcode `\$12\catcode
  `\&12\catcode `\#12\catcode `\^12\catcode `\_12\catcode `\%12\relax}%
\providecommand \@@startlink[1]{}%
\providecommand \@@endlink[0]{}%
\providecommand \url  [0]{\begingroup\@sanitize@url \@url }%
\providecommand \@url [1]{\endgroup\@href {#1}{\urlprefix }}%
\providecommand \urlprefix  [0]{URL }%
\providecommand \Eprint [0]{\href }%
\providecommand \doibase [0]{http://dx.doi.org/}%
\providecommand \selectlanguage [0]{\@gobble}%
\providecommand \bibinfo  [0]{\@secondoftwo}%
\providecommand \bibfield  [0]{\@secondoftwo}%
\providecommand \translation [1]{[#1]}%
\providecommand \BibitemOpen [0]{}%
\providecommand \bibitemStop [0]{}%
\providecommand \bibitemNoStop [0]{.\EOS\space}%
\providecommand \EOS [0]{\spacefactor3000\relax}%
\providecommand \BibitemShut  [1]{\csname bibitem#1\endcsname}%
\let\auto@bib@innerbib\@empty
\bibitem [{\citenamefont {Alexandrov}(2007)}]{polaronbook2007}%
  \BibitemOpen
  \bibinfo {editor} {\bibfnamefont {A.~S.}\ \bibnamefont {Alexandrov}},\ ed.,\
  \href@noop {} {\emph {\bibinfo {title} {Polarons in Advanced Materials}}}\
  (\bibinfo  {publisher} {Canopus Publishing and Springer Verlag GmbH},\
  \bibinfo {address} {Bristol (UK)},\ \bibinfo {year} {2007})\BibitemShut
  {NoStop}%
\bibitem [{\citenamefont {Peierls}(1979)}]{Peierls}%
  \BibitemOpen
  \bibfield  {author} {\bibinfo {author} {\bibfnamefont {R.}~\bibnamefont
  {Peierls}},\ }\href@noop {} {\emph {\bibinfo {title} {Surprises in
  Theoretical Physics}}}\ (\bibinfo  {publisher} {Princeton University Press},\
  \bibinfo {address} {New Jersey},\ \bibinfo {year} {1979})\BibitemShut
  {NoStop}%
\bibitem [{\citenamefont {Devreese}\ and\ \citenamefont
  {Alexandrov}(2009)}]{0034-4885-72-6-066501}%
  \BibitemOpen
  \bibfield  {author} {\bibinfo {author} {\bibfnamefont {J.~T.}\ \bibnamefont
  {Devreese}}\ and\ \bibinfo {author} {\bibfnamefont {A.~S.}\ \bibnamefont
  {Alexandrov}},\ }\href@noop {} {\bibfield  {journal} {\bibinfo  {journal}
  {Rep. Prog. Phys.}\ }\textbf {\bibinfo {volume} {72}},\ \bibinfo {pages}
  {066501} (\bibinfo {year} {2009})}\BibitemShut {NoStop}%
\bibitem [{\citenamefont {Chollet}\ \emph {et~al.}(2005)\citenamefont
  {Chollet}, \citenamefont {Guerin}, \citenamefont {Uchida}, \citenamefont
  {Fukaya}, \citenamefont {Shimoda}, \citenamefont {Ishikawa}, \citenamefont
  {Matsuda}, \citenamefont {Hasegawa}, \citenamefont {Ota}, \citenamefont
  {Yamochi}, \citenamefont {Saito}, \citenamefont {Tazaki}, \citenamefont
  {Adachi},\ and\ \citenamefont {Koshihara}}]{Chollet07012005}%
  \BibitemOpen
  \bibfield  {author} {\bibinfo {author} {\bibfnamefont {M.}~\bibnamefont
  {Chollet}}, \bibinfo {author} {\bibfnamefont {L.}~\bibnamefont {Guerin}},
  \bibinfo {author} {\bibfnamefont {N.}~\bibnamefont {Uchida}}, \bibinfo
  {author} {\bibfnamefont {S.}~\bibnamefont {Fukaya}}, \bibinfo {author}
  {\bibfnamefont {H.}~\bibnamefont {Shimoda}}, \bibinfo {author} {\bibfnamefont
  {T.}~\bibnamefont {Ishikawa}}, \bibinfo {author} {\bibfnamefont
  {K.}~\bibnamefont {Matsuda}}, \bibinfo {author} {\bibfnamefont
  {T.}~\bibnamefont {Hasegawa}}, \bibinfo {author} {\bibfnamefont
  {A.}~\bibnamefont {Ota}}, \bibinfo {author} {\bibfnamefont {H.}~\bibnamefont
  {Yamochi}}, \bibinfo {author} {\bibfnamefont {G.}~\bibnamefont {Saito}},
  \bibinfo {author} {\bibfnamefont {R.}~\bibnamefont {Tazaki}}, \bibinfo
  {author} {\bibfnamefont {S.-i.}\ \bibnamefont {Adachi}}, \ and\ \bibinfo
  {author} {\bibfnamefont {S.-y.}\ \bibnamefont {Koshihara}},\ }\href {\doibase
  10.1126/science.1105067} {\bibfield  {journal} {\bibinfo  {journal}
  {Science}\ }\textbf {\bibinfo {volume} {307}},\ \bibinfo {pages} {86}
  (\bibinfo {year} {2005})}\BibitemShut {NoStop}%
\bibitem [{\citenamefont {Yonemitsu}\ and\ \citenamefont
  {Nasu}(2008)}]{Yonemitsu20081}%
  \BibitemOpen
  \bibfield  {author} {\bibinfo {author} {\bibfnamefont {K.}~\bibnamefont
  {Yonemitsu}}\ and\ \bibinfo {author} {\bibfnamefont {K.}~\bibnamefont
  {Nasu}},\ }\href {\doibase http://dx.doi.org/10.1016/j.physrep.2008.04.008}
  {\bibfield  {journal} {\bibinfo  {journal} {Phys. Rep.}\ }\textbf {\bibinfo
  {volume} {465}},\ \bibinfo {pages} {1 } (\bibinfo {year} {2008})}\BibitemShut
  {NoStop}%
\bibitem [{\citenamefont {Schollw\"ock}(2005)}]{Schollwoeck05_rev}%
  \BibitemOpen
  \bibfield  {author} {\bibinfo {author} {\bibfnamefont {U.}~\bibnamefont
  {Schollw\"ock}},\ }\href@noop {} {\bibfield  {journal} {\bibinfo  {journal}
  {Rev. Mod. Phys.}\ }\textbf {\bibinfo {volume} {77}},\ \bibinfo {pages} {259}
  (\bibinfo {year} {2005})}\BibitemShut {NoStop}%
\bibitem [{\citenamefont {Bursill}\ \emph {et~al.}(1998)\citenamefont
  {Bursill}, \citenamefont {McKenzie},\ and\ \citenamefont {Hamer}}]{BuMKHa98}%
  \BibitemOpen
  \bibfield  {author} {\bibinfo {author} {\bibfnamefont {R.~J.}\ \bibnamefont
  {Bursill}}, \bibinfo {author} {\bibfnamefont {R.~H.}\ \bibnamefont
  {McKenzie}}, \ and\ \bibinfo {author} {\bibfnamefont {C.~J.}\ \bibnamefont
  {Hamer}},\ }\href@noop {} {\bibfield  {journal} {\bibinfo  {journal} {Phys.
  Rev. Lett.}\ }\textbf {\bibinfo {volume} {80}},\ \bibinfo {pages} {5607}
  (\bibinfo {year} {1998})}\BibitemShut {NoStop}%
\bibitem [{\citenamefont {Jeckelmann}\ \emph {et~al.}(1999)\citenamefont
  {Jeckelmann}, \citenamefont {Zhang},\ and\ \citenamefont {White}}]{JeZhWh99}%
  \BibitemOpen
  \bibfield  {author} {\bibinfo {author} {\bibfnamefont {E.}~\bibnamefont
  {Jeckelmann}}, \bibinfo {author} {\bibfnamefont {C.}~\bibnamefont {Zhang}}, \
  and\ \bibinfo {author} {\bibfnamefont {S.~R.}\ \bibnamefont {White}},\
  }\href@noop {} {\bibfield  {journal} {\bibinfo  {journal} {Phys. Rev. B}\
  }\textbf {\bibinfo {volume} {60}},\ \bibinfo {pages} {7950} (\bibinfo {year}
  {1999})}\BibitemShut {NoStop}%
\bibitem [{\citenamefont {Blankenbecler}\ \emph {et~al.}(1981)\citenamefont
  {Blankenbecler}, \citenamefont {Scalapino},\ and\ \citenamefont
  {Sugar}}]{BlScSu81}%
  \BibitemOpen
  \bibfield  {author} {\bibinfo {author} {\bibfnamefont {R.}~\bibnamefont
  {Blankenbecler}}, \bibinfo {author} {\bibfnamefont {D.~J.}\ \bibnamefont
  {Scalapino}}, \ and\ \bibinfo {author} {\bibfnamefont {R.~L.}\ \bibnamefont
  {Sugar}},\ }\href@noop {} {\bibfield  {journal} {\bibinfo  {journal} {Phys.
  Rev. D}\ }\textbf {\bibinfo {volume} {24}},\ \bibinfo {pages} {2278}
  (\bibinfo {year} {1981})}\BibitemShut {NoStop}%
\bibitem [{\citenamefont {{De Raedt}}\ and\ \citenamefont
  {Lagendijk}(1982)}]{dRLa82}%
  \BibitemOpen
  \bibfield  {author} {\bibinfo {author} {\bibfnamefont {H.}~\bibnamefont {{De
  Raedt}}}\ and\ \bibinfo {author} {\bibfnamefont {A.}~\bibnamefont
  {Lagendijk}},\ }\href@noop {} {\bibfield  {journal} {\bibinfo  {journal}
  {Phys. Rev. Lett.}\ }\textbf {\bibinfo {volume} {49}},\ \bibinfo {pages}
  {1522} (\bibinfo {year} {1982})}\BibitemShut {NoStop}%
\bibitem [{\citenamefont {Fradkin}\ and\ \citenamefont
  {Hirsch}(1983)}]{PhysRevB.27.1680}%
  \BibitemOpen
  \bibfield  {author} {\bibinfo {author} {\bibfnamefont {E.}~\bibnamefont
  {Fradkin}}\ and\ \bibinfo {author} {\bibfnamefont {J.~E.}\ \bibnamefont
  {Hirsch}},\ }\href {\doibase 10.1103/PhysRevB.27.1680} {\bibfield  {journal}
  {\bibinfo  {journal} {Phys. Rev. B}\ }\textbf {\bibinfo {volume} {27}},\
  \bibinfo {pages} {1680} (\bibinfo {year} {1983})}\BibitemShut {NoStop}%
\bibitem [{\citenamefont {Kornilovitch}(1998)}]{Ko98}%
  \BibitemOpen
  \bibfield  {author} {\bibinfo {author} {\bibfnamefont {P.~E.}\ \bibnamefont
  {Kornilovitch}},\ }\href@noop {} {\bibfield  {journal} {\bibinfo  {journal}
  {Phys. Rev. Lett.}\ }\textbf {\bibinfo {volume} {81}},\ \bibinfo {pages}
  {5382} (\bibinfo {year} {1998})}\BibitemShut {NoStop}%
\bibitem [{\citenamefont {Prokof'ev}\ and\ \citenamefont
  {Svistunov}(1998)}]{PrSv98}%
  \BibitemOpen
  \bibfield  {author} {\bibinfo {author} {\bibfnamefont {N.~V.}\ \bibnamefont
  {Prokof'ev}}\ and\ \bibinfo {author} {\bibfnamefont {B.~V.}\ \bibnamefont
  {Svistunov}},\ }\href@noop {} {\bibfield  {journal} {\bibinfo  {journal}
  {Phys. Rev. Lett.}\ }\textbf {\bibinfo {volume} {81}},\ \bibinfo {pages}
  {2514} (\bibinfo {year} {1998})}\BibitemShut {NoStop}%
\bibitem [{\citenamefont {Sandvik}\ and\ \citenamefont
  {Campbell}(1999)}]{PhysRevLett.83.195}%
  \BibitemOpen
  \bibfield  {author} {\bibinfo {author} {\bibfnamefont {A.~W.}\ \bibnamefont
  {Sandvik}}\ and\ \bibinfo {author} {\bibfnamefont {D.~K.}\ \bibnamefont
  {Campbell}},\ }\href {\doibase 10.1103/PhysRevLett.83.195} {\bibfield
  {journal} {\bibinfo  {journal} {Phys. Rev. Lett.}\ }\textbf {\bibinfo
  {volume} {83}},\ \bibinfo {pages} {195} (\bibinfo {year} {1999})}\BibitemShut
  {NoStop}%
\bibitem [{\citenamefont {Assaad}\ and\ \citenamefont {Lang}(2007)}]{Assaad07}%
  \BibitemOpen
  \bibfield  {author} {\bibinfo {author} {\bibfnamefont {F.~F.}\ \bibnamefont
  {Assaad}}\ and\ \bibinfo {author} {\bibfnamefont {T.~C.}\ \bibnamefont
  {Lang}},\ }\href {\doibase 10.1103/PhysRevB.76.035116} {\bibfield  {journal}
  {\bibinfo  {journal} {Phys. Rev. B}\ }\textbf {\bibinfo {volume} {76}},\
  \bibinfo {eid} {035116} (\bibinfo {year} {2007})}\BibitemShut {NoStop}%
\bibitem [{\citenamefont {Bon\v{c}a}\ \emph {et~al.}(1999)\citenamefont
  {Bon\v{c}a}, \citenamefont {Trugman},\ and\ \citenamefont
  {Batistic}}]{BoTrBa99}%
  \BibitemOpen
  \bibfield  {author} {\bibinfo {author} {\bibfnamefont {J.}~\bibnamefont
  {Bon\v{c}a}}, \bibinfo {author} {\bibfnamefont {S.~A.}\ \bibnamefont
  {Trugman}}, \ and\ \bibinfo {author} {\bibfnamefont {I.}~\bibnamefont
  {Batistic}},\ }\href@noop {} {\bibfield  {journal} {\bibinfo  {journal}
  {Phys. Rev. B}\ }\textbf {\bibinfo {volume} {60}},\ \bibinfo {pages} {1633}
  (\bibinfo {year} {1999})}\BibitemShut {NoStop}%
\bibitem [{\citenamefont {Sykora}\ \emph {et~al.}(2005)\citenamefont {Sykora},
  \citenamefont {H{\"u}bsch}, \citenamefont {Becker}, \citenamefont {Wellein},\
  and\ \citenamefont {Fehske}}]{SyHuBeWeFe04}%
  \BibitemOpen
  \bibfield  {author} {\bibinfo {author} {\bibfnamefont {S.}~\bibnamefont
  {Sykora}}, \bibinfo {author} {\bibfnamefont {A.}~\bibnamefont {H{\"u}bsch}},
  \bibinfo {author} {\bibfnamefont {K.~W.}\ \bibnamefont {Becker}}, \bibinfo
  {author} {\bibfnamefont {G.}~\bibnamefont {Wellein}}, \ and\ \bibinfo
  {author} {\bibfnamefont {H.}~\bibnamefont {Fehske}},\ }\href@noop {}
  {\bibfield  {journal} {\bibinfo  {journal} {Phys. Rev. B}\ }\textbf {\bibinfo
  {volume} {71}},\ \bibinfo {pages} {045112} (\bibinfo {year}
  {2005})}\BibitemShut {NoStop}%
\bibitem [{\citenamefont {Zhao}\ \emph {et~al.}(2005)\citenamefont {Zhao},
  \citenamefont {Wu},\ and\ \citenamefont {Lin}}]{ZhWuLi05}%
  \BibitemOpen
  \bibfield  {author} {\bibinfo {author} {\bibfnamefont {H.}~\bibnamefont
  {Zhao}}, \bibinfo {author} {\bibfnamefont {C.~Q.}\ \bibnamefont {Wu}}, \ and\
  \bibinfo {author} {\bibfnamefont {H.~Q.}\ \bibnamefont {Lin}},\ }\href@noop
  {} {\bibfield  {journal} {\bibinfo  {journal} {Phys. Rev. B}\ }\textbf
  {\bibinfo {volume} {71}},\ \bibinfo {pages} {115201} (\bibinfo {year}
  {2005})}\BibitemShut {NoStop}%
\bibitem [{\citenamefont {Hohenadler}\ \emph {et~al.}(2006)\citenamefont
  {Hohenadler}, \citenamefont {Wellein}, \citenamefont {Bishop}, \citenamefont
  {Alvermann},\ and\ \citenamefont {Fehske}}]{Hohenadler06}%
  \BibitemOpen
  \bibfield  {author} {\bibinfo {author} {\bibfnamefont {M.}~\bibnamefont
  {Hohenadler}}, \bibinfo {author} {\bibfnamefont {G.}~\bibnamefont {Wellein}},
  \bibinfo {author} {\bibfnamefont {A.~R.}\ \bibnamefont {Bishop}}, \bibinfo
  {author} {\bibfnamefont {A.}~\bibnamefont {Alvermann}}, \ and\ \bibinfo
  {author} {\bibfnamefont {H.}~\bibnamefont {Fehske}},\ }\href@noop {}
  {\bibfield  {journal} {\bibinfo  {journal} {Phys. Rev. B}\ }\textbf {\bibinfo
  {volume} {73}},\ \bibinfo {eid} {245120} (\bibinfo {year}
  {2006})}\BibitemShut {NoStop}%
\bibitem [{\citenamefont {Voit}\ \emph {et~al.}(2000)\citenamefont {Voit},
  \citenamefont {Perfetti}, \citenamefont {Zwick}, \citenamefont {Berger},
  \citenamefont {Margaritondo}, \citenamefont {Gr\"uner}, \citenamefont
  {H\"ochst},\ and\ \citenamefont {Grioni}}]{Vo.Pe.Zw.Be.Ma.Gr.Ho.Gr.00}%
  \BibitemOpen
  \bibfield  {author} {\bibinfo {author} {\bibfnamefont {J.}~\bibnamefont
  {Voit}}, \bibinfo {author} {\bibfnamefont {L.}~\bibnamefont {Perfetti}},
  \bibinfo {author} {\bibfnamefont {F.}~\bibnamefont {Zwick}}, \bibinfo
  {author} {\bibfnamefont {H.}~\bibnamefont {Berger}}, \bibinfo {author}
  {\bibfnamefont {G.}~\bibnamefont {Margaritondo}}, \bibinfo {author}
  {\bibfnamefont {G.}~\bibnamefont {Gr\"uner}}, \bibinfo {author}
  {\bibfnamefont {H.}~\bibnamefont {H\"ochst}}, \ and\ \bibinfo {author}
  {\bibfnamefont {M.}~\bibnamefont {Grioni}},\ }\href@noop {} {\bibfield
  {journal} {\bibinfo  {journal} {Science}\ }\textbf {\bibinfo {volume}
  {290}},\ \bibinfo {pages} {501} (\bibinfo {year} {2000})}\BibitemShut
  {NoStop}%
\bibitem [{\citenamefont {Hohenadler}\ \emph {et~al.}(2011)\citenamefont
  {Hohenadler}, \citenamefont {Fehske},\ and\ \citenamefont
  {Assaad}}]{Hohenadler10a}%
  \BibitemOpen
  \bibfield  {author} {\bibinfo {author} {\bibfnamefont {M.}~\bibnamefont
  {Hohenadler}}, \bibinfo {author} {\bibfnamefont {H.}~\bibnamefont {Fehske}},
  \ and\ \bibinfo {author} {\bibfnamefont {F.~F.}\ \bibnamefont {Assaad}},\
  }\href {\doibase 10.1103/PhysRevB.83.115105} {\bibfield  {journal} {\bibinfo
  {journal} {Phys. Rev. B}\ }\textbf {\bibinfo {volume} {83}},\ \bibinfo
  {pages} {115105} (\bibinfo {year} {2011})}\BibitemShut {NoStop}%
\bibitem [{\citenamefont {Creffield}\ \emph {et~al.}(2005)\citenamefont
  {Creffield}, \citenamefont {Sangiovanni},\ and\ \citenamefont
  {Capone}}]{CrSaCa05}%
  \BibitemOpen
  \bibfield  {author} {\bibinfo {author} {\bibfnamefont {C.~E.}\ \bibnamefont
  {Creffield}}, \bibinfo {author} {\bibfnamefont {G.}~\bibnamefont
  {Sangiovanni}}, \ and\ \bibinfo {author} {\bibfnamefont {M.}~\bibnamefont
  {Capone}},\ }\href@noop {} {\bibfield  {journal} {\bibinfo  {journal} {Eur.
  Phys. J. B}\ }\textbf {\bibinfo {volume} {44}},\ \bibinfo {pages} {175}
  (\bibinfo {year} {2005})}\BibitemShut {NoStop}%
\bibitem [{\citenamefont {Sykora}\ \emph
  {et~al.}(2006{\natexlab{a}})\citenamefont {Sykora}, \citenamefont
  {H\"ubsch},\ and\ \citenamefont {Becker}}]{Sykora06}%
  \BibitemOpen
  \bibfield  {author} {\bibinfo {author} {\bibfnamefont {S.}~\bibnamefont
  {Sykora}}, \bibinfo {author} {\bibfnamefont {A.}~\bibnamefont {H\"ubsch}}, \
  and\ \bibinfo {author} {\bibfnamefont {K.~W.}\ \bibnamefont {Becker}},\
  }\href@noop {} {\bibfield  {journal} {\bibinfo  {journal} {Europhys. Lett.}\
  }\textbf {\bibinfo {volume} {76}},\ \bibinfo {pages} {644} (\bibinfo {year}
  {2006}{\natexlab{a}})}\BibitemShut {NoStop}%
\bibitem [{\citenamefont {Loos}\ \emph {et~al.}(2006)\citenamefont {Loos},
  \citenamefont {Hohenadler}, \citenamefont {Alvermann},\ and\ \citenamefont
  {Fehske}}]{LoHoAlFe06}%
  \BibitemOpen
  \bibfield  {author} {\bibinfo {author} {\bibfnamefont {J.}~\bibnamefont
  {Loos}}, \bibinfo {author} {\bibfnamefont {M.}~\bibnamefont {Hohenadler}},
  \bibinfo {author} {\bibfnamefont {A.}~\bibnamefont {Alvermann}}, \ and\
  \bibinfo {author} {\bibfnamefont {H.}~\bibnamefont {Fehske}},\ }\href@noop {}
  {\bibfield  {journal} {\bibinfo  {journal} {J. Phys.: Condens. Matter}\
  }\textbf {\bibinfo {volume} {18}},\ \bibinfo {pages} {7299} (\bibinfo {year}
  {2006})}\BibitemShut {NoStop}%
\bibitem [{\citenamefont {Sykora}\ \emph
  {et~al.}(2006{\natexlab{b}})\citenamefont {Sykora}, \citenamefont
  {H\"{u}bsch},\ and\ \citenamefont {Becker}}]{SyHuBe05}%
  \BibitemOpen
  \bibfield  {author} {\bibinfo {author} {\bibfnamefont {S.}~\bibnamefont
  {Sykora}}, \bibinfo {author} {\bibfnamefont {A.}~\bibnamefont {H\"{u}bsch}},
  \ and\ \bibinfo {author} {\bibfnamefont {K.~W.}\ \bibnamefont {Becker}},\
  }\href@noop {} {\bibfield  {journal} {\bibinfo  {journal} {Eur. Phys. J. B}\
  }\textbf {\bibinfo {volume} {51}},\ \bibinfo {pages} {181} (\bibinfo {year}
  {2006}{\natexlab{b}})}\BibitemShut {NoStop}%
\bibitem [{\citenamefont {Fehske}\ \emph {et~al.}(2004)\citenamefont {Fehske},
  \citenamefont {Wellein}, \citenamefont {Hager}, \citenamefont {Wei{\ss}e},\
  and\ \citenamefont {Bishop}}]{FeWeHaWeBi03}%
  \BibitemOpen
  \bibfield  {author} {\bibinfo {author} {\bibfnamefont {H.}~\bibnamefont
  {Fehske}}, \bibinfo {author} {\bibfnamefont {G.}~\bibnamefont {Wellein}},
  \bibinfo {author} {\bibfnamefont {G.}~\bibnamefont {Hager}}, \bibinfo
  {author} {\bibfnamefont {A.}~\bibnamefont {Wei{\ss}e}}, \ and\ \bibinfo
  {author} {\bibfnamefont {A.~R.}\ \bibnamefont {Bishop}},\ }\href@noop {}
  {\bibfield  {journal} {\bibinfo  {journal} {Phys. Rev. B}\ }\textbf {\bibinfo
  {volume} {69}},\ \bibinfo {pages} {165115} (\bibinfo {year}
  {2004})}\BibitemShut {NoStop}%
\bibitem [{\citenamefont {Payeur}\ and\ \citenamefont
  {S\'en\'echal}(2011)}]{PhysRevB.83.033104}%
  \BibitemOpen
  \bibfield  {author} {\bibinfo {author} {\bibfnamefont {A.}~\bibnamefont
  {Payeur}}\ and\ \bibinfo {author} {\bibfnamefont {D.}~\bibnamefont
  {S\'en\'echal}},\ }\href {\doibase 10.1103/PhysRevB.83.033104} {\bibfield
  {journal} {\bibinfo  {journal} {Phys. Rev. B}\ }\textbf {\bibinfo {volume}
  {83}},\ \bibinfo {pages} {033104} (\bibinfo {year} {2011})}\BibitemShut
  {NoStop}%
\bibitem [{\citenamefont {Hohenadler}\ and\ \citenamefont
  {Assaad}(2013)}]{PhysRevB87.075149}%
  \BibitemOpen
  \bibfield  {author} {\bibinfo {author} {\bibfnamefont {M.}~\bibnamefont
  {Hohenadler}}\ and\ \bibinfo {author} {\bibfnamefont {F.~F.}\ \bibnamefont
  {Assaad}},\ }\href@noop {} {\bibfield  {journal} {\bibinfo  {journal} {Phys.
  Rev. B}\ }\textbf {\bibinfo {volume} {87}},\ \bibinfo {pages} {075149}
  (\bibinfo {year} {2013})}\BibitemShut {NoStop}%
\bibitem [{\citenamefont {Ning}\ \emph {et~al.}(2006)\citenamefont {Ning},
  \citenamefont {Zhao}, \citenamefont {Wu},\ and\ \citenamefont
  {Lin}}]{NiZhWuLi05}%
  \BibitemOpen
  \bibfield  {author} {\bibinfo {author} {\bibfnamefont {W.~Q.}\ \bibnamefont
  {Ning}}, \bibinfo {author} {\bibfnamefont {H.}~\bibnamefont {Zhao}}, \bibinfo
  {author} {\bibfnamefont {C.~Q.}\ \bibnamefont {Wu}}, \ and\ \bibinfo {author}
  {\bibfnamefont {H.~Q.}\ \bibnamefont {Lin}},\ }\href@noop {} {\bibfield
  {journal} {\bibinfo  {journal} {Phys. Rev. Lett.}\ }\textbf {\bibinfo
  {volume} {96}},\ \bibinfo {pages} {156402} (\bibinfo {year}
  {2006})}\BibitemShut {NoStop}%
\bibitem [{\citenamefont {Matsueda}\ \emph {et~al.}(2006)\citenamefont
  {Matsueda}, \citenamefont {Tohyama},\ and\ \citenamefont
  {Maekawa}}]{MaToMa05}%
  \BibitemOpen
  \bibfield  {author} {\bibinfo {author} {\bibfnamefont {H.}~\bibnamefont
  {Matsueda}}, \bibinfo {author} {\bibfnamefont {T.}~\bibnamefont {Tohyama}}, \
  and\ \bibinfo {author} {\bibfnamefont {S.}~\bibnamefont {Maekawa}},\ }\href
  {\doibase 10.1103/PhysRevB.74.241103} {\bibfield  {journal} {\bibinfo
  {journal} {Phys. Rev. B}\ }\textbf {\bibinfo {volume} {74}},\ \bibinfo
  {pages} {241103} (\bibinfo {year} {2006})}\BibitemShut {NoStop}%
\bibitem [{\citenamefont {Assaad}(2008)}]{Assaad08}%
  \BibitemOpen
  \bibfield  {author} {\bibinfo {author} {\bibfnamefont {F.~F.}\ \bibnamefont
  {Assaad}},\ }\href {\doibase 10.1103/PhysRevB.78.155124} {\bibfield
  {journal} {\bibinfo  {journal} {Phys. Rev. B}\ }\textbf {\bibinfo {volume}
  {78}},\ \bibinfo {eid} {155124} (\bibinfo {year} {2008})}\BibitemShut
  {NoStop}%
\bibitem [{\citenamefont {Hohenadler}\ \emph {et~al.}(2012)\citenamefont
  {Hohenadler}, \citenamefont {Assaad},\ and\ \citenamefont
  {Fehske}}]{Ho.As.Fe.12}%
  \BibitemOpen
  \bibfield  {author} {\bibinfo {author} {\bibfnamefont {M.}~\bibnamefont
  {Hohenadler}}, \bibinfo {author} {\bibfnamefont {F.~F.}\ \bibnamefont
  {Assaad}}, \ and\ \bibinfo {author} {\bibfnamefont {H.}~\bibnamefont
  {Fehske}},\ }\href@noop {} {\bibfield  {journal} {\bibinfo  {journal} {Phys.
  Rev. Lett.}\ }\textbf {\bibinfo {volume} {109}},\ \bibinfo {pages} {116407}
  (\bibinfo {year} {2012})}\BibitemShut {NoStop}%
\bibitem [{\citenamefont {Hohenadler}(2013)}]{PhysRevB.88.064303}%
  \BibitemOpen
  \bibfield  {author} {\bibinfo {author} {\bibfnamefont {M.}~\bibnamefont
  {Hohenadler}},\ }\href {\doibase 10.1103/PhysRevB.88.064303} {\bibfield
  {journal} {\bibinfo  {journal} {Phys. Rev. B}\ }\textbf {\bibinfo {volume}
  {88}},\ \bibinfo {pages} {064303} (\bibinfo {year} {2013})}\BibitemShut
  {NoStop}%
\bibitem [{\citenamefont {Weber}\ \emph {et~al.}(2015)\citenamefont {Weber},
  \citenamefont {Assaad},\ and\ \citenamefont {Hohenadler}}]{WeAsHo15_I}%
  \BibitemOpen
  \bibfield  {author} {\bibinfo {author} {\bibfnamefont {M.}~\bibnamefont
  {Weber}}, \bibinfo {author} {\bibfnamefont {F.~F.}\ \bibnamefont {Assaad}}, \
  and\ \bibinfo {author} {\bibfnamefont {M.}~\bibnamefont {Hohenadler}},\
  }\href@noop {} {\bibfield  {journal} {\bibinfo  {journal} {Phys. Rev. B}\ }
  \textbf {\bibinfo {volume} {91}},\ \bibinfo {pages} {245147} (\bibinfo {year} {2015})}\BibitemShut
  {NoStop}%
\bibitem [{\citenamefont {Holstein}(1959)}]{Ho59a}%
  \BibitemOpen
  \bibfield  {author} {\bibinfo {author} {\bibfnamefont {T.}~\bibnamefont
  {Holstein}},\ }\href@noop {} {\bibfield  {journal} {\bibinfo  {journal} {Ann.
  Phys. (N.Y.)}\ }\textbf {\bibinfo {volume} {8}},\ \bibinfo {pages} {325
  (1959); {\bf 8}, 343} (\bibinfo {year} {1959})}\BibitemShut {NoStop}%
\bibitem [{\citenamefont {Clay}\ and\ \citenamefont {Hardikar}(2005)}]{ClHa05}%
  \BibitemOpen
  \bibfield  {author} {\bibinfo {author} {\bibfnamefont {R.~T.}\ \bibnamefont
  {Clay}}\ and\ \bibinfo {author} {\bibfnamefont {R.~P.}\ \bibnamefont
  {Hardikar}},\ }\href@noop {} {\bibfield  {journal} {\bibinfo  {journal}
  {Phys. Rev. Lett.}\ }\textbf {\bibinfo {volume} {95}},\ \bibinfo {pages}
  {096401} (\bibinfo {year} {2005})}\BibitemShut {NoStop}%
\bibitem [{\citenamefont {Hardikar}\ and\ \citenamefont
  {Clay}(2007)}]{hardikar:245103}%
  \BibitemOpen
  \bibfield  {author} {\bibinfo {author} {\bibfnamefont {R.~P.}\ \bibnamefont
  {Hardikar}}\ and\ \bibinfo {author} {\bibfnamefont {R.~T.}\ \bibnamefont
  {Clay}},\ }\href {\doibase 10.1103/PhysRevB.75.245103} {\bibfield  {journal}
  {\bibinfo  {journal} {Phys. Rev. B}\ }\textbf {\bibinfo {volume} {75}},\
  \bibinfo {eid} {245103} (\bibinfo {year} {2007})}\BibitemShut {NoStop}%
\bibitem [{\citenamefont {Fehske}\ \emph {et~al.}(2008)\citenamefont {Fehske},
  \citenamefont {Hager},\ and\ \citenamefont
  {Jeckelmann}}]{0295-5075-84-5-57001}%
  \BibitemOpen
  \bibfield  {author} {\bibinfo {author} {\bibfnamefont {H.}~\bibnamefont
  {Fehske}}, \bibinfo {author} {\bibfnamefont {G.}~\bibnamefont {Hager}}, \
  and\ \bibinfo {author} {\bibfnamefont {E.}~\bibnamefont {Jeckelmann}},\
  }\href@noop {} {\bibfield  {journal} {\bibinfo  {journal} {Europhys. Lett.}\
  }\textbf {\bibinfo {volume} {84}},\ \bibinfo {pages} {57001} (\bibinfo {year}
  {2008})}\BibitemShut {NoStop}%
\bibitem [{\citenamefont {Bakrim}\ and\ \citenamefont
  {Bourbonnais}(2015)}]{Bakrim2015}%
  \BibitemOpen
  \bibfield  {author} {\bibinfo {author} {\bibfnamefont {H.}~\bibnamefont
  {Bakrim}}\ and\ \bibinfo {author} {\bibfnamefont {C.}~\bibnamefont
  {Bourbonnais}},\ }\href {\doibase 10.1103/PhysRevB.91.085114} {\bibfield
  {journal} {\bibinfo  {journal} {Phys. Rev. B}\ }\textbf {\bibinfo {volume}
  {91}},\ \bibinfo {pages} {085114} (\bibinfo {year} {2015})}\BibitemShut
  {NoStop}%
\bibitem [{\citenamefont {Rubtsov}\ \emph {et~al.}(2005)\citenamefont
  {Rubtsov}, \citenamefont {Savkin},\ and\ \citenamefont
  {Lichtenstein}}]{Rubtsov05}%
  \BibitemOpen
  \bibfield  {author} {\bibinfo {author} {\bibfnamefont {A.~N.}\ \bibnamefont
  {Rubtsov}}, \bibinfo {author} {\bibfnamefont {V.~V.}\ \bibnamefont {Savkin}},
  \ and\ \bibinfo {author} {\bibfnamefont {A.~I.}\ \bibnamefont
  {Lichtenstein}},\ }\href@noop {} {\bibfield  {journal} {\bibinfo  {journal}
  {Phys. Rev. B}\ }\textbf {\bibinfo {volume} {72}},\ \bibinfo {pages} {035122}
  (\bibinfo {year} {2005})}\BibitemShut {NoStop}%
\bibitem [{\citenamefont {Feynman}(1955)}]{Feynman55}%
  \BibitemOpen
  \bibfield  {author} {\bibinfo {author} {\bibfnamefont {R.~P.}\ \bibnamefont
  {Feynman}},\ }\href@noop {} {\bibfield  {journal} {\bibinfo  {journal} {Phys.
  Rev.}\ }\textbf {\bibinfo {volume} {97}},\ \bibinfo {pages} {660} (\bibinfo
  {year} {1955})}\BibitemShut {NoStop}%
\bibitem [{\citenamefont {Werner}\ and\ \citenamefont
  {Millis}(2007)}]{werner:146404}%
  \BibitemOpen
  \bibfield  {author} {\bibinfo {author} {\bibfnamefont {P.}~\bibnamefont
  {Werner}}\ and\ \bibinfo {author} {\bibfnamefont {A.~J.}\ \bibnamefont
  {Millis}},\ }\href@noop {} {\bibfield  {journal} {\bibinfo  {journal} {Phys.
  Rev. Lett.}\ }\textbf {\bibinfo {volume} {99}},\ \bibinfo {eid} {146404}
  (\bibinfo {year} {2007})}\BibitemShut {NoStop}%
\bibitem [{col()}]{colorscheme}%
  \BibitemOpen
  \href@noop {} {}\bibinfo {howpublished} {A Schneider et al.,
  Gnuplot-colorbrewer: ColorBrewer color schemes for gnuplot. Zenodo.
  10.5281/zenodo.10282.}\BibitemShut {Stop}%
\bibitem [{\citenamefont {Hewson}\ and\ \citenamefont
  {Meyer}(2002)}]{0953-8984-14-3-312}%
  \BibitemOpen
  \bibfield  {author} {\bibinfo {author} {\bibfnamefont {A.~C.}\ \bibnamefont
  {Hewson}}\ and\ \bibinfo {author} {\bibfnamefont {D.}~\bibnamefont {Meyer}},\
  }\href {http://stacks.iop.org/0953-8984/14/i=3/a=312} {\bibfield  {journal}
  {\bibinfo  {journal} {J. Phys.: Condens. Matter}\ }\textbf {\bibinfo {volume}
  {14}},\ \bibinfo {pages} {427} (\bibinfo {year} {2002})}\BibitemShut
  {NoStop}%
\bibitem [{\citenamefont {Beach}(2004)}]{Beach04a}%
  \BibitemOpen
  \bibfield  {author} {\bibinfo {author} {\bibfnamefont {K.~S.~D.}\
  \bibnamefont {Beach}},\ }\href@noop {} {\bibfield  {journal} {\bibinfo
  {journal} {arXiv:0403055}\ } (\bibinfo {year} {2004})}\BibitemShut {NoStop}%
\bibitem [{\citenamefont {Engelsberg}\ and\ \citenamefont
  {Varga}(1964)}]{PhysRev.136.A1582}%
  \BibitemOpen
  \bibfield  {author} {\bibinfo {author} {\bibfnamefont {S.}~\bibnamefont
  {Engelsberg}}\ and\ \bibinfo {author} {\bibfnamefont {B.~B.}\ \bibnamefont
  {Varga}},\ }\href {\doibase 10.1103/PhysRev.136.A1582} {\bibfield  {journal}
  {\bibinfo  {journal} {Phys. Rev.}\ }\textbf {\bibinfo {volume} {136}},\
  \bibinfo {pages} {A1582} (\bibinfo {year} {1964})}\BibitemShut {NoStop}%
\bibitem [{\citenamefont {Michel}\ and\ \citenamefont
  {Evertz}(2007)}]{Michel07}%
  \BibitemOpen
  \bibfield  {author} {\bibinfo {author} {\bibfnamefont {F.}~\bibnamefont
  {Michel}}\ and\ \bibinfo {author} {\bibfnamefont {H.~G.}\ \bibnamefont
  {Evertz}},\ }\href@noop {} {\bibfield  {journal} {\bibinfo  {journal}
  {arXiv:0705.0799}\ }}\BibitemShut {NoStop}%
\bibitem [{\citenamefont {Migdal}(1958)}]{Migdal58}%
  \BibitemOpen
  \bibfield  {author} {\bibinfo {author} {\bibfnamefont {A.}~\bibnamefont
  {Migdal}},\ }\href@noop {} {\bibfield  {journal} {\bibinfo  {journal}
  {J. Exp. Theor. Phys.}\
  }\textbf {\bibinfo {volume} {34}},\ \bibinfo {pages} {996} (\bibinfo {year}
  {1958})}\BibitemShut {NoStop}%
\bibitem [{\citenamefont {Alexandrov}\ \emph {et~al.}(1994)\citenamefont
  {Alexandrov}, \citenamefont {Kabanov},\ and\ \citenamefont {Ray}}]{AlKaRa94}%
  \BibitemOpen
  \bibfield  {author} {\bibinfo {author} {\bibfnamefont {A.~S.}\ \bibnamefont
  {Alexandrov}}, \bibinfo {author} {\bibfnamefont {V.~V.}\ \bibnamefont
  {Kabanov}}, \ and\ \bibinfo {author} {\bibfnamefont {D.~K.}\ \bibnamefont
  {Ray}},\ }\href@noop {} {\bibfield  {journal} {\bibinfo  {journal} {Phys.
  Rev. B}\ }\textbf {\bibinfo {volume} {49}},\ \bibinfo {pages} {9915}
  (\bibinfo {year} {1994})}\BibitemShut {NoStop}%
\bibitem [{Note1()}]{Note1}%
  \BibitemOpen
  \bibinfo {note} {Our field theory implies a continuum limit such that
  momentum conservation is exact. Hence the identification of $2k_\protect
  \text {F}$ to $-2k_\protect \text {F}$ for the half-filled band is not
  justified.}\BibitemShut {Stop}%
\bibitem [{\citenamefont {Lee}\ \emph {et~al.}(1974)\citenamefont {Lee},
  \citenamefont {Rice},\ and\ \citenamefont {Anderson}}]{Lee74}%
  \BibitemOpen
  \bibfield  {author} {\bibinfo {author} {\bibfnamefont {P.}~\bibnamefont
  {Lee}}, \bibinfo {author} {\bibfnamefont {T.}~\bibnamefont {Rice}}, \ and\
  \bibinfo {author} {\bibfnamefont {P.}~\bibnamefont {Anderson}},\ }\href@noop
  {} {\bibfield  {journal} {\bibinfo  {journal} {Solid State Commun.}\ }\textbf
  {\bibinfo {volume} {14}},\ \bibinfo {pages} {703} (\bibinfo {year}
  {1974})}\BibitemShut {NoStop}%
\bibitem [{\citenamefont {Roy}\ \emph {et~al.}(2013)\citenamefont {Roy},
  \citenamefont {Juri\ifmmode \check{c}\else \v{c}\fi{}i\ifmmode~\acute{c}\else
  \'{c}\fi{}},\ and\ \citenamefont {Herbut}}]{PhysRevB.87.041401}%
  \BibitemOpen
  \bibfield  {author} {\bibinfo {author} {\bibfnamefont {B.}~\bibnamefont
  {Roy}}, \bibinfo {author} {\bibfnamefont {V.}~\bibnamefont {Juri\ifmmode
  \check{c}\else \v{c}\fi{}i\ifmmode~\acute{c}\else \'{c}\fi{}}}, \ and\
  \bibinfo {author} {\bibfnamefont {I.~F.}\ \bibnamefont {Herbut}},\ }\href
  {\doibase 10.1103/PhysRevB.87.041401} {\bibfield  {journal} {\bibinfo
  {journal} {Phys. Rev. B}\ }\textbf {\bibinfo {volume} {87}},\ \bibinfo
  {pages} {041401} (\bibinfo {year} {2013})}\BibitemShut {NoStop}%
\bibitem [{\citenamefont {Lang}\ \emph {et~al.}(2013)\citenamefont {Lang},
  \citenamefont {Meng}, \citenamefont {Muramatsu}, \citenamefont {Wessel},\
  and\ \citenamefont {Assaad}}]{PhysRevLett.111.066401}%
  \BibitemOpen
  \bibfield  {author} {\bibinfo {author} {\bibfnamefont {T.~C.}\ \bibnamefont
  {Lang}}, \bibinfo {author} {\bibfnamefont {Z.~Y.}\ \bibnamefont {Meng}},
  \bibinfo {author} {\bibfnamefont {A.}~\bibnamefont {Muramatsu}}, \bibinfo
  {author} {\bibfnamefont {S.}~\bibnamefont {Wessel}}, \ and\ \bibinfo {author}
  {\bibfnamefont {F.~F.}\ \bibnamefont {Assaad}},\ }\href {\doibase
  10.1103/PhysRevLett.111.066401} {\bibfield  {journal} {\bibinfo  {journal}
  {Phys. Rev. Lett.}\ }\textbf {\bibinfo {volume} {111}},\ \bibinfo {pages}
  {066401} (\bibinfo {year} {2013})}\BibitemShut {NoStop}%
\bibitem [{Note2()}]{Note2}%
  \BibitemOpen
  \bibinfo {note} {We consider finite lattices with an arbitrary cutoff for the
  phonons. The same results can be obtained by evaluating $\protect
  \mathaccentV {bar}016{B}(q,\omega )=\protect \operatorname {Im}[D(q,\omega +
  i\eta )]/\pi $ for the general case, where $D(q,z)$ has a branch cut on the
  real axis.}\BibitemShut {Stop}%
\bibitem [{Note3()}]{Note3}%
  \BibitemOpen
  \bibinfo {note} {In general, this need not be the case.}\BibitemShut {Stop}%
\bibitem [{Note4()}]{Note4}%
  \BibitemOpen
  \bibinfo {note} {However, the integrated weight $\DOTSI \intop \ilimits@ B(q,\omega ) d\omega $ may
  be very different for $\omega <\omega _0$ and $\omega >\omega
  _0$.}\BibitemShut {Stop}%
\end{thebibliography}

%

\end{document}